\documentclass[a4paper,12pt]{article}
\usepackage{color}
\usepackage{amssymb}
\usepackage{amsmath}
\usepackage{graphicx}
\usepackage{hyperref}
\title{\bf Pinning down neutrino oscillation parameters in the 2--3
sector with a mgnetised atmospheric neutrino detector: a new study} 

\author{Lakshmi~S.~Mohan$^{1,2}$ and D.~Indumathi$^1$\\
$^{1}${The Institute of Mathematical Sciences, Chennai 600 113, India}\\
$^{2}${Physical Research Laboratory, Ahmedabad 380 009, India}\\}

\date{\today} 

\begin{document}
\maketitle

\begin{abstract}
 We determine the sensitivity to neutrino oscillation
parameters from a study of atmospheric neutrinos in a magnetised
detector such as the ICAL at the proposed India-based Neutrino
Observatory. In such a detector that can {\em separately} count
$\nu_\mu$ and $\overline{\nu}_\mu$-induced events, the relatively
smaller (about 5\%) uncertainties on the neutrino--anti-neutrino flux
ratios translate to a constraint in the $\chi^2$ analysis that results
in a significant improvement in the precision with which neutrino
oscillation parameters such as $\sin^2\theta_{23}$ can be determined.
Such an effect is unique to all magnetisable detectors and constitutes a
great advantage in determining neutrino oscillation parameters using
such detectors. Such a study has been performed for the first time here.
Along with an increase in the kinematic range compared to earlier
analyses, this results in sensitivities to oscillation parameters in the
2--3 sector that are comparable to or better than those from accelerator
experiments where the fluxes are significantly higher. For example, the $1\sigma$
precisions on $\sin^2\theta_{23}$ and $|\Delta{m^2_{32(31)}}|$ achievable
for 500 kTon yr exposure of ICAL are $\sim9\%$ and $\sim2.5\%$
respectively for both normal and inverted hierarchies. The mass hierarchy
sensitivity achievable with this combination when the true hierarchy
is normal (inverted) for the same exposure is $\Delta\chi^2\approx8.5$
($\Delta\chi^2\approx9.5$).
\end{abstract}

\section{Introduction}\label{intro}
One of the open questions in neutrino physics is the mass ordering of the neutrinos; whether 
they are ordered normally or inverted. Many experiments intend to
determine the mass ordering, 
of which the 50 kton Iron Calorimeter (ICAL) detector at the proposed India-based Neutrino 
observatory is one ambitious experiment \cite{WP}. 
ICAL will be a magnetised iron calorimeter mainly sensitive to muons produced in
the charged-current (CC) interactions of atmospheric muon neutrinos (and
anti-neutrinos) with the iron target in the detector. It can distinguish
CC muon-neutrino-induced events from anti-neutrino-induced ones since
the former interaction produces $\mu^-$ while the latter produces $\mu^+$
in the detector and ICAL has excellent muon charge identification (cid)
capability. This is also crucial to determine precisely the momentum
of the muons through bending in the magnetic field. Since matter
effects are different between neutrino and anti-neutrino propagation
in the Earth, this feature can help resolve the neutrino mass ordering
by determining the sign of the 2--3 mass-squared difference $\Delta
m^2_{32} \equiv m^2_3-m^2_2$, $m_i,~i=1,2,3$, being the neutrino mass
eigenstates \cite{imsc-hie-paper}. In addition, the matter effects
improve the sensitivity to the magnitude $\vert \Delta m^2_{32}
\vert$ of this mass-squared difference as well as to the 2--3
mixing angle, $\theta_{23}$, provided the across-generation mixing
angle $\theta_{13}$ is rather well-known, which is indeed the case
\cite{double-chooz1,dayabay1,dayabay2,reno-results,double-chooz-results}.

Many previous analyses have been reported, projecting the sensitivities of ICAL detector to oscillation 
parameters in the 2--3 sector \cite{TApre,TAhie,3dMMD,WP} as also the
mass ordering. The sensitivity to mass hierarchy is 
directly proportional to the value of $\theta_{13}$ which is quite
precisely known \cite{dayabay-2015,dayabay-2016,reno-2016,double-chooz-2014,double-chooz-2015,nu-fit}.
It also depends on the ability of ICAL to separate neutrino and
anti-neutrino events which is possible since ICAL is magnetised. While
there is an uncertainty of about 20\% on the atmospheric neutrino fluxes
themselves, the uncertainty on their {\em ratios} is much smaller, about
5\%, and was ignored in earlier analyses \cite{TApre,TAhie,3dMMD}. In
this paper, we show that this smaller uncertainty on the ratio acts as a constraint that in
turn significantly shrinks the allowed parameter space, especially for
$\sin^2\theta_{23}$. For instance, we will see that the precision on 
$\sin^2\theta_{23}$ decreases from 13\% to 9\% in a certain analysis mode
when this constraint is included. This is generally true for all
magnetised detectors. To our knowledge such an effect has not been
discussed in the literature earlier.

The paper is structured as follows. All the results
in this paper are with detailed simulation studies of the physics processes at the ICAL detector. The main steps 
 involved in this are neutrino event generation, inclusion of detector responses and efficiencies, inclusion of 
 oscillations, binning in observables and $\chi^2$ analysis. The procedure of neutrino event generation with the 
 NUANCE neutrino \cite{nuance} generator and the implementation of oscillations are discussed in detail in 
 Section~\ref{nu-evt}. The choice of observables and kinematic regions used in the analysis, along with the 
 inclusion of detector responses are discussed in detail in
 Section~\ref{obs}. The effect of increasing the energy 
 range of observed muons is also explained in this section. The detailed $\chi^2$ analysis and a discussion of 
 the systematic errors that have been considered are presented in Section~\ref{chisq}. 
 The results of precision measurements and hierarchy sensitivity studies are shown in Section~\ref{results}.
 The impact of the additional pull in the $\nu/\overline{\nu}$ flux
 ratio implemented in this analysis is discussed in detail in
 Section~\ref{eff-11-pull}.
 The summary and conclusions are given in Section~\ref{summary}.
\section{Neutrino events generation}\label{nu-evt}
The interactions of interest in ICAL are the CC interactions of
$\nu_\mu$ and $\overline{\nu}_\mu$ with the iron target in ICAL. These
$\nu_\mu$ ($\overline{\nu}_\mu$) in ICAL come from both $\nu_\mu$
and $\nu_e$ atmospheric fluxes via $\nu_\mu\rightarrow\nu_\mu$ and
$\nu_e\rightarrow\nu_\mu$ oscillations. The first channel gives the number
of $\nu_\mu$ events which have survived and the second, subdominant,
one gives the number from oscillations of $\nu_e$ to $\nu_\mu$. The
number of events ICAL sees will be a sum of these events. Thus,
\begin{eqnarray} \nonumber
\frac{d^2N}{d E_\mu d\cos\theta_\mu} & = & t \times {n_d}\times
\int{d E_\nu d\cos\theta_\nu d\phi_\nu} \times \\ 
& & \hspace{0.5cm}
\left[P_{\mu\mu} \frac{d^3\Phi_\mu}{d E_\nu d\cos\theta_\nu d\phi_\nu}+
P_{e\mu} \frac{d^3\Phi_e}{d E_\nu d\cos\theta_\nu d\phi_\nu}
\right] \times
\frac{d\sigma_\mu (E_\nu)}{d E_\mu d\cos\theta_\mu}~,
\label{toteve}
\end{eqnarray}
where $n_d$ is the number of target nucleons in the detector, $\sigma_\mu$
is the differential neutrino interaction cross section in terms of the
energy and direction of the CC lepton produced, $\Phi_\mu$ and $\Phi_e$ are the
$\nu_\mu$ and $\nu_e$ fluxes and $P_{\alpha\beta}$ is the oscillation
probability of $\nu_\alpha\rightarrow\nu_\beta$.

The number of {\em unoscillated} events over an exposure time $t$ in a
bin of ($E_\mu,\cos\theta_\mu$) is obtained from the NUANCE neutrino
generator using the Honda 3D atmospheric neutrino fluxes
\cite{honda-paper}, neutrino-nucleus cross-sections, and a simplified
ICAL detector geometry. While NUANCE lists details of all the final state
particles including the muon and all hadrons, ICAL will be optimised to
determine accurately the energy and direction of the muons (seen as a
clean track in the detector) and the summed energy of all the hadrons
in the final state (since it cannot distinguish individual hadrons).

Even though the analyses are done for a smaller number of years (say
10), a huge sample of NUANCE events for a very large number of years
(here 1000 years) is generated and scaled down to the required number
of years during the analysis. This is mainly done to reduce the effect
of statistical (Monte Carlo) fluctuations on sensitivity studies,
which may alter the results. A detailed discussion about the effect of
fluctuations on oscillation sensitivity studies will be discussed in
Appendix~\ref{fluct}.

A sample of 1000 years of unoscillated events was generated using
NUANCE. Two sets were generated:
\begin{enumerate}
\item CC muon events using the $\Phi_\mu$ flux and 
\item CC muon events obtained by swapping
$\Phi_e\leftrightarrow\Phi_\mu$ fluxes.
\end{enumerate}
This generates the so-called muon- and swapped-muon events that correspond to the two terms in Eq.~\ref{toteve}.

\subsection{Oscillation probabilities}
\label{oscprob}

These events are oscillated depending on the neutrino oscillation
parameters being used. The oscillation probabilities are calculated by
considering the full three flavour oscillations in the presence of matter
effects. The Preliminary Reference Earth Model (PREM) profile \cite{prem}
has been used to model the varying Earth matter densities encountered by
the neutrinos during their travel through the Earth. The Runge-Kutta
solver method is used to calculate the oscillation probabilities
\cite{imsc-par-paper} for various energies $E_\nu$ and distances $L$,
or equivalently, $\cos\theta_\nu$ ($\theta_\nu$ being the zenith angle) 
of the neutrino. Further discussion of the oscillation probabilities and
plots of a few sample curves are presented in the next section after listing 
the kinematical range of interest.

The oscillation is applied event by event (for both muon and swapped
muon events) as discussed in detail in Ref.~\cite{3dMMD} and it is a time consuming 
process since the actual sample contains 1000 years of events. The central values of 
the oscillations parameters are given in Table~\ref{osc-par-3sig} along with their known
$3\sigma$ range. Note that $\delta_{CP}$ is currently unknown and its
true value has been assumed to be $0^\circ$ for the
purposes of this calculation. Furthermore, since ICAL is insensitive
\cite{TAhie} to this parameter, it has been kept fixed in the
calculation, along with the values of the 1--2 oscillation parameters
$\Delta m^2_{21}$ and $\sin^2 \theta_{12}$ which also do not affect the
results.

\begin{table}[htp]
\centering 
\begin{tabular}{|c|c|c|}
\hline
Parameter & True value & Marginalization range \\
\hline
$\theta_{13}$ & 8.729$^\circ$ & [7.671$^\circ$, 9.685$^\circ$] \\
$\sin^{2}\theta_{23}$ & 0.5 & [0.36, 0.66] \\
$\Delta{m^2_{\rm eff}}$ & $\pm2.4\times10^{-3}~{\rm eV}^2$ & [2.1, 2.6]$\times10^{-3}~{\rm eV}^2$ (NH) \\
  &	& [-2.6, -2.1]$\times10^{-3}~{\rm eV}^2$ (IH) \\
$\sin^{2}\theta_{12}$ & 0.304 & Not marginalised \\
$\Delta{m^{2}_{21}}$ & $7.6\times10^{-5}~{\rm eV}^2$ & Not marginalised \\
$\delta_{CP}$ & 0$^\circ$ & Not marginalised \\
\hline  
\end{tabular}
\caption{Main oscillation parameters used in the current analysis. In the second column are the true values of these parameters used to simulate the 
``observed'' data set. True value is the value at which the data is simulated. More details are given in the main text. For precision measurement of 
each parameter, all others are varied except that parameter in the analysis.}
\label{osc-par-3sig}
\end{table}

It is convenient to define the effective mass-squared difference
$\Delta m^2_{\rm eff}$ which is the measured quantity whose value is
related to $\Delta m^2_{31}$ and $\Delta m^2_{21}$ as \cite{andre,nuno}:
\begin{equation}
\Delta m_{\rm eff}^2 = \Delta m_{31}^2-\Delta m_{21}^2
\left(\cos^2\theta_{12}-\cos\delta_{CP}\sin\theta_{13}
\sin 2\theta_{12} \tan\theta_{23}\right)~.
\label{dmeff2}
\end{equation}
When $\Delta m^{2}_{\rm eff}$ is varied within its 3$\sigma$ range,
the mass-squared differences are determined according to
\begin{eqnarray} \nonumber
\Delta m^2_{31} & = & \Delta m^2_{\rm eff}+\Delta m_{21}^2
\left(\cos^2\theta_{12}-\cos\delta_{CP}\sin\theta_{13}
\sin2\theta_{12}\tan\theta_{23}\right)~; \\
\Delta m^2_{32} & = & \Delta m^2_{31}-\Delta m^2_{21}~,
\end{eqnarray}
for normal hierarchy when $\Delta{m}_{\rm eff}^2 > 0$, with
$\Delta m^2_{31} \leftrightarrow -\Delta m^2_{32}$ for inverted
hierarchy when $\Delta{m}_{\rm eff}^2 < 0$. A neater definition of the
mass ordering can be obtained by defining the quantity
\begin{eqnarray}
\Delta m^2 \equiv m_3^2 - \frac{(m_1+m_2)^2}{2} & = & \Delta m^2_{32} + \frac{1}{2} \Delta m^2_{21}~.
\end{eqnarray}
Then, switching the ordering from normal to inverted is exactly
equivalent to the interchange $\Delta m^2 \leftrightarrow - \Delta m^2$,
with no change in its magnitude. However, since the marginalisation is
to be done on the observed quantity $\Delta m^2_{\rm eff}$, we use this
quantity, but need to keep in mind that $\Delta m^2_{32} \leftrightarrow
-\Delta m^2_{31}$ when the ordering is flipped between NH and IH in this
case.

\section{Choice of Observables and Kinematic Regions}\label{obs}
The expression in Eq.~\ref{toteve} is for the ideal case when the detector
has perfect resolutions and 100\% efficiencies. In this analysis,
realistic resolutions and efficiencies obtained from GEANT-4
based simulation studies of ICAL \cite{mupaper1,peripheralmu,mthesis,hres1,hrest} have been incorporated; this
not only reduces the overall events due to the reconstruction efficiency
factor but also smears out the final state (or observed) muon energy
and direction and that of the hadron energy as well.

\subsection{ICAL detection efficiencies}\label{eff}
Detailed simulations analyses of the reconstruction efficiency, direction
and energy resolution of muons in ICAL have been presented in
Refs.~\cite{mupaper1,peripheralmu}. In addition, the relative 
cid (charge identification) efficiency of muons (ability of ICAL to distinguish $\mu^-$ from
$\mu^+$) has also been presented here. The detector has good direction reconstruction capability 
(better than about $1^\circ$ for few-GeV muons) and excellent cid efficiency (better than 99\% 
for few-GeV muons) also for muons. The detailed simulation studies of the response of ICAL to hadrons have been
presented in Refs.~\cite{hres1,hrest}. Hadron hits are identified and
calibrated to reconstruct the energy of hadrons in neutrino-induced
interactions in ICAL. The present analysis has used these results to
simulate the observed events in ICAL. Note that the efficiency of an
event is taken to be the ability to see a muon, that is, to be able to
reconstruct it. Hence when hadron energy is added as the third observable,
the efficiency in detecting an event remains the same.

At the time this calculation was begun, the responses of both muons
and hadrons in the peripheral parts of ICAL was not completely
understood. Hence, instead of propagating the NUANCE events
through the simulated ICAL detector in GEANT and obtaining a more
realistic set of ``observed'' values of the energy and momentum of
the final state particles, the true values of these variables were
smeared according to the resolutions obtained in the earlier studies
\cite{mupaper1,hres1}.

It should be noted that instead of reconstructing the neutrino energy
and direction using the muon and hadron information and then binning in
neutrino energy, the analyses have been done by taking all the observables
separately. This is because of the poor energy and direction resolution
of neutrinos in ICAL detector owing to the fact that they are driven by
the responses of the detector to hadrons, which are worse compared to
those of muons. Still, the addition of the extra information regarding
hadrons improves the sensitivity of ICAL to oscillation parameters, as
shown in Ref.~\cite{3dMMD}.

\subsection{Effect of extending the energy range of observed muons}\label{ext}
The first highlight of this paper is widening the energy range of the 
observables, especially that of observed muons. Since ICAL is optimised for 
muon detection, it is desirable to make use of all the events available to perform 
the oscillation analysis. As opposed to all the earlier studies in ICAL \cite{TApre,TAhie,3dMMD,WP}
which restricted themselves to performing the analyses in the energy range of only 
1--11 GeV of the observed muon energy, the analysis we present here uses the 
region of the observed muon energy $E^{obs}_\mu$ = 0.5--25 GeV. It will be seen 
in Section~\ref{results} the inclusion of the higher energy bins beyond
the upper limit of 11 GeV used in earlier studies, improves the results.

\begin{figure}[htb]
\centering
\includegraphics[width=0.49\textwidth]{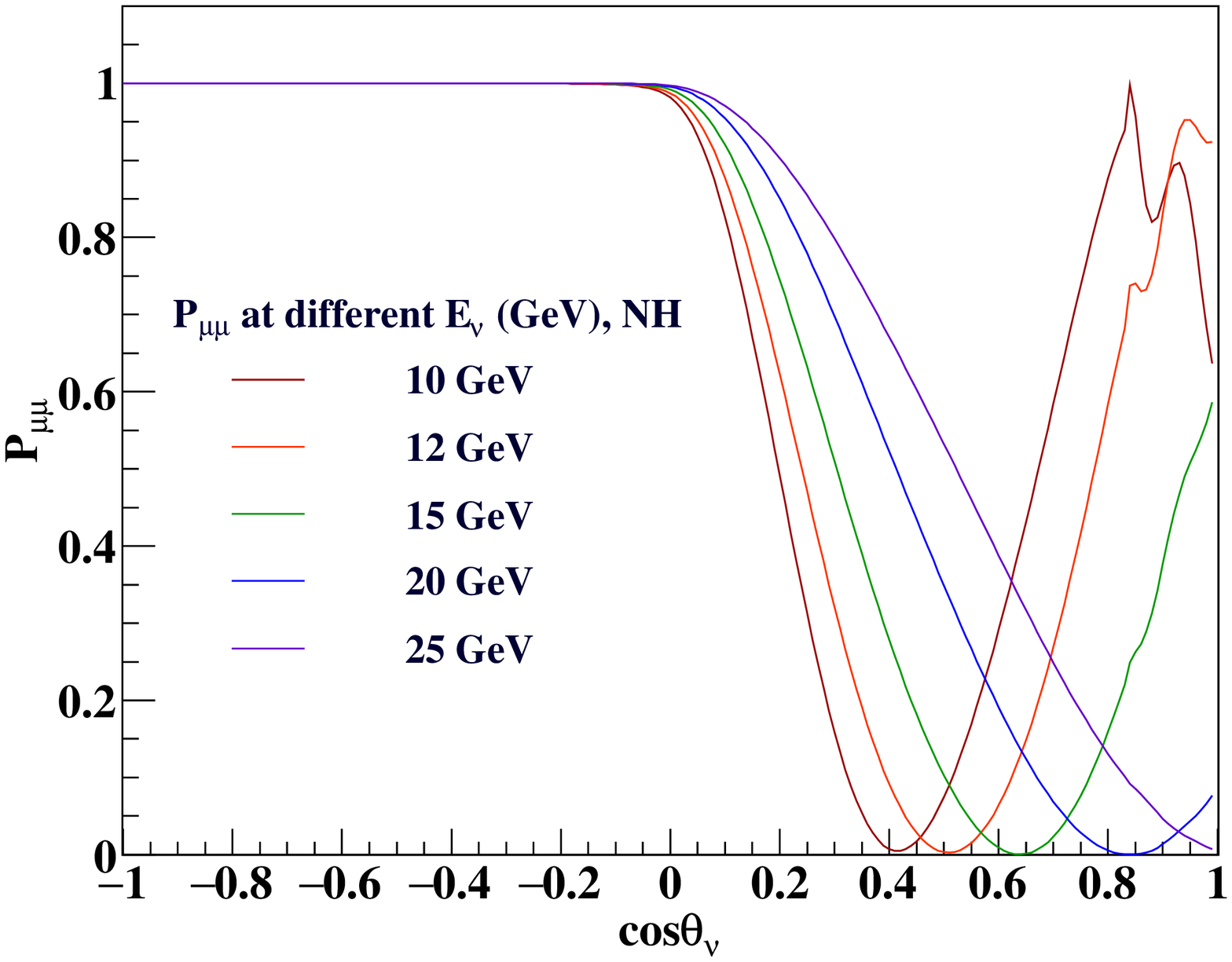}
\hfill
\includegraphics[width=0.49\textwidth]{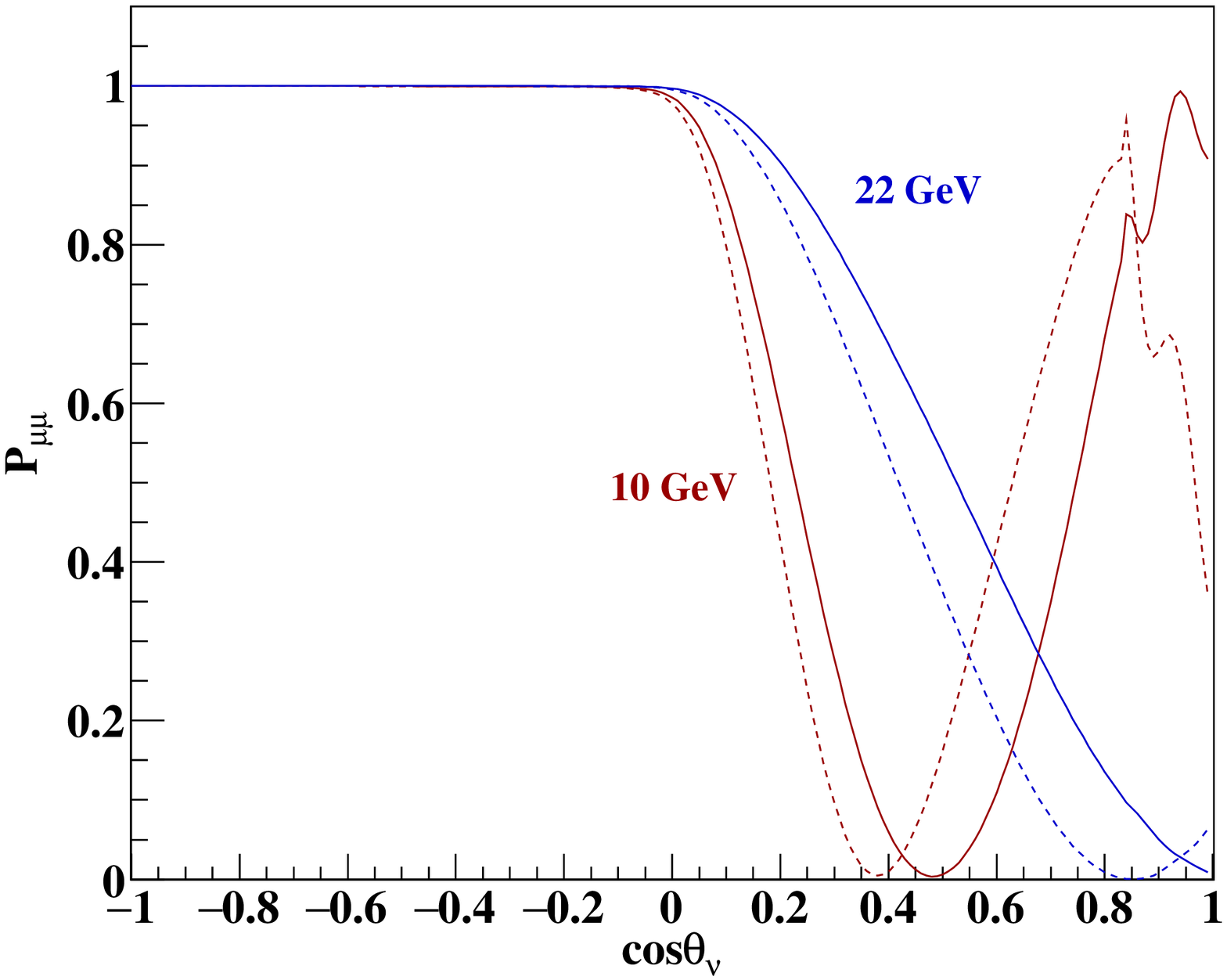}
\caption{Left: Oscillation probability $P_{\mu\mu}$ as a function of zenith
angle for different values of neutrino energy, $E_\nu = 10,12,15,20,25$ GeV
assuming true hierarchy as normal. Right: The same probability shown as solid 
(dashed) lines for two different
energy values for $\Delta m^2_{\rm eff} = 2.1~(2.6) \times 10^{-3}$ eV$^2$
to show the sensitivity to this parameter at higher energies.}
\label{fig:pmumu}
\end{figure}

The motivation to use the extended range of the observed muon
energy is seen in Fig.~\ref{fig:pmumu} where the
dominant oscillation probability, $P_{\mu\mu}$, is shown as a 
function of the zenith angle $\cos\theta$
for different values of the neutrino energy, $E_\nu \ge 10$ GeV. With
increase in energy, the curve smooths out (matter effects become
small so that $P_{\mu\mu} \sim \overline{P}_{\mu\mu}$) and 
correspondingly $P_{e\mu}$ becomes vanishingly small. Note also the
vanishing of $P_{\mu\mu}$ for high energies, $E_\nu \gtrsim 20$ GeV,
in the upward direction ($\cos\theta \to 1$).

The sensitivity to $\Delta m^2_{\rm eff}$ is shown for 
two different energies, $E_\nu = 10, 22$ GeV, (representing the last 
energy bins of the previous analysis and the
present one respectively) in Fig.~\ref{fig:pmumu} as well.
The minimum moves to the left with increasing
$\Delta m^2_{\rm eff}$ so that the solid (dashed) line corresponds to
$\Delta m^2_{\rm eff} = 2.1~(2.6) \times 10^{-3}$ eV$^2$, which is the 
presently allowed $3\sigma$ range. It can be
seen that the position of the minimum of $P_{\mu\mu}$ is more sensitive
to the value of $\Delta m^2_{\rm eff}$ at the larger value of energy,
although the probability itself is not sensitive to the {\em sign} of this
quantity at this energy. Hence the inclusion of the higher energy bins
improves the sensitivity to these oscillation parameters, as we shall see.

\subsection{The binning scheme}\label{bin}
This is similar to, and an extension of, the one used in the earlier
analyses \cite{3dMMD}.
The observables in the analysis are the observed (i.e., smeared) muon
energy $E^{obs}_\mu$, observed muon direction $\cos\theta^{obs}_\mu$
and observed hadron energy $E'^{obs}_{had}$, where the true total hadron
energy is defined as \cite{3dMMD} $E'_{had} \equiv E_\nu - E_\mu$. There
are two different analysis sets, one in which only the muon energy
and direction, $(E^{obs}_\mu, \cos\theta^{obs}_\mu)$, are used, called
the 2D (mu only) binning scheme and the other in which all the three
observables $(E^{obs}_\mu, \cos\theta^{obs}_\mu, E'^{obs}_{had})$ are
used, which is also known as the 3D (or with-hadron) binning scheme. The
details of the two binning schemes are shown in Table~\ref{mo-wh-bins}.

\begin{table}[htb]
\centering
\begin{tabular}{|c|c|c|c|}
\hline
Observable & Range & Bin width & No.of bins \\ 
\hline
& [0.5, 4] & 0.5 & 7 \\
& [4, 7] & 1 & 3 \\
$E^{obs}_{\mu}$ (GeV)& [7, 11] & 4 & 1 \\
{\color{blue} (15 bins)} & [11, 12.5] & 1.5 & 1 \\
& [12.5, 15] & 2.5 & 1 \\
& [15, 25] & 5 & 2 \\
\hline
& [-1.0, 0.0] & 0.2 & 5 \\
$\cos\theta^{obs}_{\mu}$ & [0.0, 0.4] & 0.10 & 4 \\
{\color{blue} (21 bins)} & [0.4, 1.0] & 0.05 & 12 \\
\hline
& [0, 2] & 1 & 2 \\
$E'^{obs}_{had}$ (GeV) & [2, 4] & 2 & 1\\
{\color{blue} (4 bins)} & [4, 15] & 11 & 1 \\
\hline
\end{tabular}
\caption{Bins of the three observables, muon energy and direction and
hadron energy, used in the analysis.}
\label{mo-wh-bins}
\end{table} 

It should be noted that in the current analysis the direction
$\cos\theta^{obs}_\mu = +1$ is taken as the up direction. Since
atmospheric neutrino oscillations are mainly in the up direction,
more bins are assigned in this region than in the down direction. The
$E^{obs}_\mu$ bins upto 1--11 GeV are taken to be same as those used in 
Ref.~\cite{3dMMD}. A bin of width 0.5 GeV is added in the lower energy range.
In the higher range of $E^{obs}_\mu$, four bins are added, two of them with 
bin width each of 1.5 GeV and 2.5 GeV respectively and the last two bins of 
width 5 GeV, thus making the total number of bins in $E^{obs}_\mu$ 15. 
The $\cos\theta^{obs}_\mu$ bins and the $E'^{obs}_{had}$ are kept the same as 
in Ref.~\cite{3dMMD}. Thus same bins are used in the overlapping energy range,
and the additional bin sizes were optimised to obtain reasonable event rate as well as
sensitivity to oscillation parameters. This gives an optimised result on the whole.

As mentioned above, the number of hadron bins was retained as before, as also the energy 
range. No extension of hadron energies beyond $E'^{obs}_{had} = 15$ GeV was used, since 
this gave only a marginal improvement in $\chi^2$ while ICAL's sensitivity to hadrons
at higher energies in terms of the number of hits in the detector tends
to saturate \cite{hres1}.

\subsection{Number of events}\label{nevt}
The true number of oscillated events is given by: 
\begin{eqnarray} \nonumber
N_{\mu^-} & = & N_{\mu^-}^{0}\times P_{\mu\mu} +
N_{e^-}^{0}\times P_{e\mu}~, \\ \nonumber
N_{\mu^+} & = & N_{\mu^+}^{0}\times \overline{P}_{\mu\mu} +
N_{e^+}^{0}\times \overline{P}_{e\mu}~,
\end{eqnarray}
where $P_{\alpha\beta}$ is the oscillation probability of a flavour
$\alpha$ to a flavour $\beta$, $N_{\mu^{\pm}}^0$ and $N_{e^{\pm}}^0$
refer to unoscillated muon and swapped-muon events generated by NUANCE
arising from survived atmospheric $\nu_\mu \to \nu_\mu$ and oscillated
atmospheric $\nu_e \to \nu_\mu$ interactions in ICAL corresponding to
the two terms in Eq.~\ref{toteve}.

The number of events per bin including the charge misidentified ones
is given as,
\begin{eqnarray} 
N^{tot}_{\mu^-}(E^{obs}_\mu,\cos\theta^{obs}_\mu) & = &
N_{\mu^-} \epsilon_{rec} \epsilon_{cid} +
N_{\mu^+} \epsilon_{rec} (1-\epsilon_{cid})~, \\ \nonumber 
N^{tot}_{\mu^+}(E^{obs}_\mu,\cos\theta^{obs}_\mu) & = &
N_{\mu^+} \epsilon_{rec} \epsilon_{cid} + 
N_{\mu^-} \epsilon_{rec} (1-\epsilon_{cid})~,
\end{eqnarray}
where $N^{tot}_{\mu^-}$ ($N^{tot}_{\mu^+}$) is the total number
of oscillated $\nu_\mu$ ($\overline{\nu}_\mu$) CC muon neutrino 
events observed in the bin $(E^{obs}_\mu,\cos\theta^{obs}_\mu)$. The quantity
$\epsilon_{rec}$ is the reconstruction efficiency of muons with a
given energy and direction and $\epsilon_{cid}$ is the relative charge
identification efficiency of the same. The reconstruction and charge
identification efficiencies for $\mu^-$ and $\mu^+$ have been taken to
be the same; studies show \cite{mupaper1} that they are only marginally
different in a few energy-$\cos\theta$ bins. Finally, the events in a bin
are considered non-zero if there is at least one event in that bin.

Now this 1000 year sample, oscillated according to the central values of
the oscillation parameters listed in Table~\ref{osc-par-3sig}, is scaled
to the required number of years to generate the ``data''. The current
precision analysis is done for 10 years of exposure of 50 kton ICAL
(500 kton year). In order to generate the ``theory'' for comparison
with ``data'' for the $\chi^2$ analysis, the oscillation parameters
are changed within their $3\sigma$ ranges and the aforementioned
processes are repeated. Different theories are generated by changing
the oscillation parameters.

\section{$\chi^2$ analysis}\label{chisq}
Systematic uncertainties play a very important role in determining the 
sensitivity to oscillation parameters in any experiment. The inclusion of 
these uncertaines always gives a worse $\chi^2$ than the 
one obtained when we have no uncertaines at all. 

In this new analysis, an extra systematic uncertainty compared to the older analyses
  is included: the uncertainty on the neutrino--antineutrino flux ratio. This has been considered for the
first time in such an analysis and will be seen to have a great impact because ICAL is a magnetised detector 
that can separate $\mu^-$ and $\mu^+$ events. With the inclusion of this uncertainty, the $\chi^2$ can no longer
be expressed as a sum of the separate contributions of neutrino and anti-neutrino
events. When the systematic errors are implemented using the usual method of pulls
\cite{kameda,ishitsuka,maltoni,kamland,sbnufact} we have, 
\begin{eqnarray} 
\chi^2_{11} & = & {\stackrel{\hbox{min}}{\displaystyle \xi_l^{\pm},\xi_6}} 
\sum^{N_{E^{obs}_{\mu}}}_{i=1}\sum^{N_{\cos\theta^{obs}_{\mu}}}_{j=1}
\left(\sum^{N_{E'^{obs}_{had}}}_{k=1}\right) 2\left[ \left(T^{+}_{ij(k)}
- D^{+}_{ij(k)} \right) - D^{+}_{ij(k)} \ln \left( 
\frac{T^{+}_{ij(k)}}{D^{+}_{ij(k)}} \right) \right] + \nonumber \\
& & 2\left[\left(T^{-}_{ij(k)} - D^{-}_{ij(k)}\right) - D^{-}_{ij(k)}
\ln\left(\frac{T^{-}_{ij(k)}}{D^{-}_{ij(k)}}\right)\right] + 
\sum^{5}_{l^{+}=1} \xi^{2}_{l^{+}} + \sum^{5}_{l^{-}=1}
\xi^{2}_{l^{-}} + \xi^{2}_6~,
\label{chisq-11p}
\end{eqnarray}
where $i,j,k$ sum over muon energy, muon angle and hadron energy bins. The number of 
theory (expected) events in each bin, with systematic errors, is given by, 
\begin{eqnarray}
T^{+}_{ij(k)} & = & T^{0+}_{ij(k)}\left(1+\sum^{5}_{l^{+}=1} 
\pi^{l^{+}}_{ij(k)}\xi_{l^{+}}+\pi_6\xi_6\right)~, \\ \nonumber
T^{-}_{ij(k)} &= & T^{0-}_{ij(k)}\left(1+\sum^{5}_{l^{-}=1}
\pi^{l^{-}}_{ij(k)}\xi_{l^{-}}-\pi_6\xi_6\right)~,
\label{td-pi6xi6}
\end{eqnarray}
where $T^{0\pm}_{ij(k)}$ is the corresponding number of events without systematic errors, 
$D^{\pm}_{ij(k)}$ is the number of ``data'' (observed) events in each
bin, and
$\xi_{l^{\pm}}$ are the pulls corresponding to the same five systematic
uncertainties, $l=1,\ldots,5$, for each of neutrino and anti-neutrino
contributions, as considered in the earlier analyses by the INO collaboration
\cite{WP,TApre,TAhie,3dMMD}.

The five systematic uncertaines include the flux normalisation, shape 
(spectral or energy dependence) uncertainty or ``tilt'' and zenith angle uncertainties, the 
cross-section unertainties, and an overall systematic uncertainty due 
to the detector response (see Ref.~\cite{3dMMD} for details).

The cross-section uncertainty is assumed to be process independent. At
high energies, the cross-section is dominated by deep inelastic
scattering (DIS) where the uncertainties are smaller; however, in the
energy range of interest here, all processes (quasi-elastic (QE),
resonance (RES) and DIS) have significant contributions and so a common
cross-section uncertainty is used.

Here the values of $\pi^l$ are taken to be the same for neutrino events and anti-neutrino events; i.e, 
$\pi^{l\pm} \equiv \pi^l; l = 1, \ldots, 5$. The values used
in this analysis are the same as those used in the earlier analysis by
the INO collaboration \cite{TApre,TAhie,3dMMD,WP}:
\begin{enumerate}
\item $\pi_1=20$\% flux normalisation error,
\item $\pi_2=10$\% cross section error,
\item $\pi_3=5$\% tilt error,
\item $\pi_4=5$\% zenith angle error, 
\item $\pi_5=5$\% overall systematics.
\end{enumerate}
In Eq. \ref{chisq-11p} $\xi_6$ is the 11th (additional) pull and $\pi_6$
is taken to be 2.5\%. The effect of the new pull can be understood by
considering its contribution
alone on the ratio of neutrino to anti-neutrino events:
\begin{eqnarray} 
\frac{N^+}{N^-} & \simeq & \frac{T^{0+}}{T^{0-}}
\frac{(1+\pi_6\xi_6)}{(1-\pi_6\xi_6)} \\ 
& \simeq & \frac{T^{0+}}{T^{0-}}(1+2\pi_6\xi_6)~.
\label{pi6-xi6}
\end{eqnarray}
This pull therefore accounts for the uncertainty in the flux ratio;
$2\pi_6$ corresponds to the 1$\sigma$ error (when $\xi_6=1$); this gives the
1$\sigma$ error on the ratio to be 5\%, consistent with
Ref.~\cite{honda-paper,honda-gaisser}. In the earlier analysis with 10
pulls only, the pulls for $N^-$ and $N^+$ were independent so that
they could be in the same or opposite directions. The introduction
of the 11th pull constrains the ratio and results in a (negative)
correlation between the normalisations of the $T^+$ and $T^-$ events,
as will become clear from the discussions presented in
Sections~\ref{results} and \ref{eff-11-pull}. Without this pull, the total 
$\chi^2$ can be simply expressed as a sum over the $\mu^-$ and $\mu^+$
contributions : 
\begin{equation}
\chi^2_{10} = \chi^2_{+}+\chi^2_{-}.
\label{chisq-10p}
\end{equation}

Note that the observed muon events have
contributions from both the $\Phi_\mu$ and $\Phi_e$ fluxes; here we have
assumed the same systematic error on both the
$\overline{\Phi}_\mu/\Phi_\mu$ and $\overline{\Phi}_e/\Phi_e$ ratios
(although, in principle, this can be included separately). We have also
ignored the small differences due to additional possible uncertainty in
the $\Phi_\mu/\Phi_e$ flux ratios since the contribution from the second
term in Eq.~\ref{toteve}, i.e., from $\nu_e \to \nu_\mu$ oscillations is
subdominant/small.

An 8\% prior at 1$\sigma$ is also added on $\sin^22\theta_{13}$, since
this quantity is known to this accuracy \cite{dayabay1,dayabay2}. No
prior is imposed on $\theta_{23}$ and $\Delta{m_{32}^2}$, since the
precision measurements of these parameters are to be carried out with
ICAL. The contribution to $\chi^2$ due to prior is defined as :
\begin{equation}
\chi_{\rm prior}^2 = \left(\frac{\sin^22\theta_{13}-
\sin^22\theta_{13}^{\rm true}}
{\sigma(\sin^22\theta_{13})}\right)^2~,
\label{chi2-prior}
\end{equation}
where $\sigma(\sin^22\theta_{13})$ = $0.08\times\sin^2\theta_{13}^{\rm true}$. 
Thus the total $\chi^2$ is defined as:
\begin{equation}
\chi_{\rm ICAL}^{2} = \chi^2+\chi_{\rm prior}^2~,
\label{chi2-ical}
\end{equation}
where $\chi^2$ corresponds suitably to $\chi^2_{11}$ (new analysis) or $\chi^2_{10}$
(repeat of the older analysis with extended energy range).

During $\chi^2$ minimisation, $\chi_{\rm ICAL}^{2}$ is first
minimised with respect to the pull variables $\xi_l$ for a given set
of oscillation parameters, then marginalised over the ranges of the
oscillation parameters $\sin^2\theta_{23}$, $\Delta m^2_{\rm eff}$
and $\sin^22\theta_{13}$ given in Table~\ref{osc-par-3sig}. The third
column of the table shows the 3$\sigma$ range over which the parameter
values are varied. These along with the best fit values of
$\theta_{12}$ and $\Delta{m_{21}^2}$ are obtained from the global fits
in Refs.~\cite{nufitpage,nufit-2012,forero,gonzalezgarcia,capozzi}. As mentioned
earlier, the parameter $\delta_{CP}$ is kept fixed at zero throughout
this analysis.

The relative precision achieved on a parameter $\lambda$ (here $\lambda$
being $\sin^2\theta_{23}$ or $|\Delta{m^2_{\rm eff}}|$) at 1$\sigma$
is expressed as :
\begin{equation}
p(\lambda) = \frac{\lambda_{\hbox{max-}2\sigma}-
\lambda_{\hbox{min-}2\sigma}}{4\lambda_{true}}~,
\label{sig1-pre}
\end{equation}
where $\lambda_{\hbox{max-}2\sigma}$ and $\lambda_{\hbox{min-}2\sigma}$
are the maximum and minimum allowed values of $\lambda$ at 2$\sigma$;
$\lambda_{true}$ is the true choice.

The statistical significance of the obtained result is denoted by
$n\sigma$, where, $n=\sqrt{\Delta\chi^2}$, which is given by:
\begin{equation}
\Delta\chi^2(\lambda) = \chi^2_{ICAL}(\lambda)-\chi^2_0~,
\end{equation}
$\chi^2_0$ being the minimum value of $\chi^2_{ICAL}$ in the allowed
parameter range. With no statistical fluctuations, $\chi^2_0 = 0$.

\section{Results: precision measurement of parameters}\label{results}

The precision measurement of oscillation parameters in the atmospheric
sector in the energy range 0.5--25 GeV is probed in the current
analysis using 500 kton year exposure of ICAL detector. Comparisons
with previous analyses in the 1--11 GeV energy range are also
done to illustrate how much better is the sensitivity with our new analysis. 
Sensitivities to the oscillation parameters $\theta_{23}$
and $|\Delta{m_{\rm eff}^2}|$ are found separately, when the other
parameter and $\theta_{13}$ are marginalised over their 3$\sigma$
ranges. Marginalisation is also done over the two possible mass
hierarchies; since atmospheric neutrino events in ICAL are sensitive
to the mass hierarchy (also discussed below), the wrong hierarchy
always gives a worse value of $\chi^2$ during marginalisation.
Typically, normal hierarchy (NH) is taken to be the true hierarchy (a couple 
of results with inverted hierarchy (IH) are shown for completeness) and
500 kton years of exposure is used (10 years of running the experiment).

\subsection{Precision measurement of $\sin^2\theta_{23}$}\label{pre-stt23}
\begin{figure}[htb]
\centering
\includegraphics[width=0.49\textwidth=0.5]{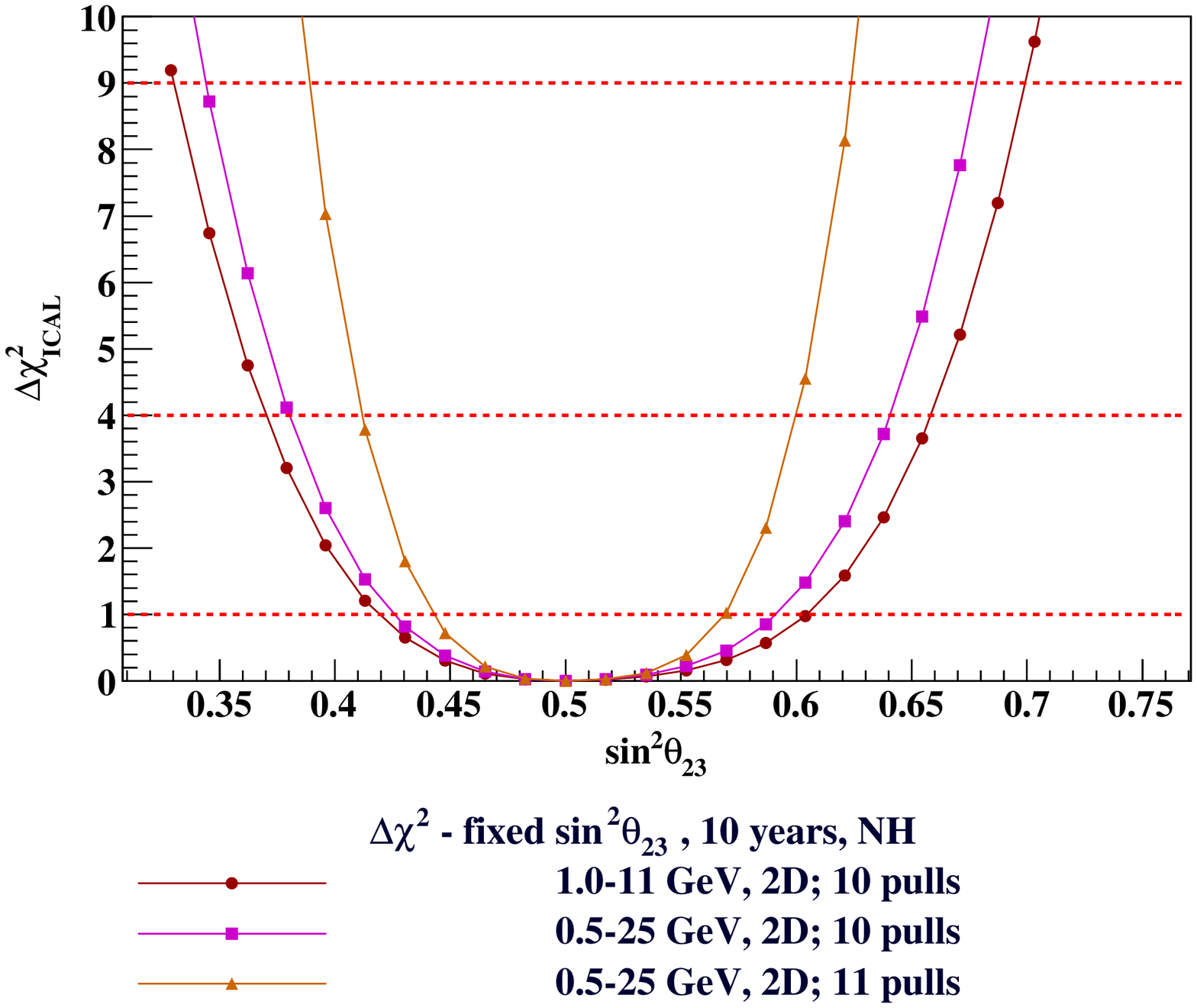}
\hfill
\includegraphics[width=0.49\textwidth]{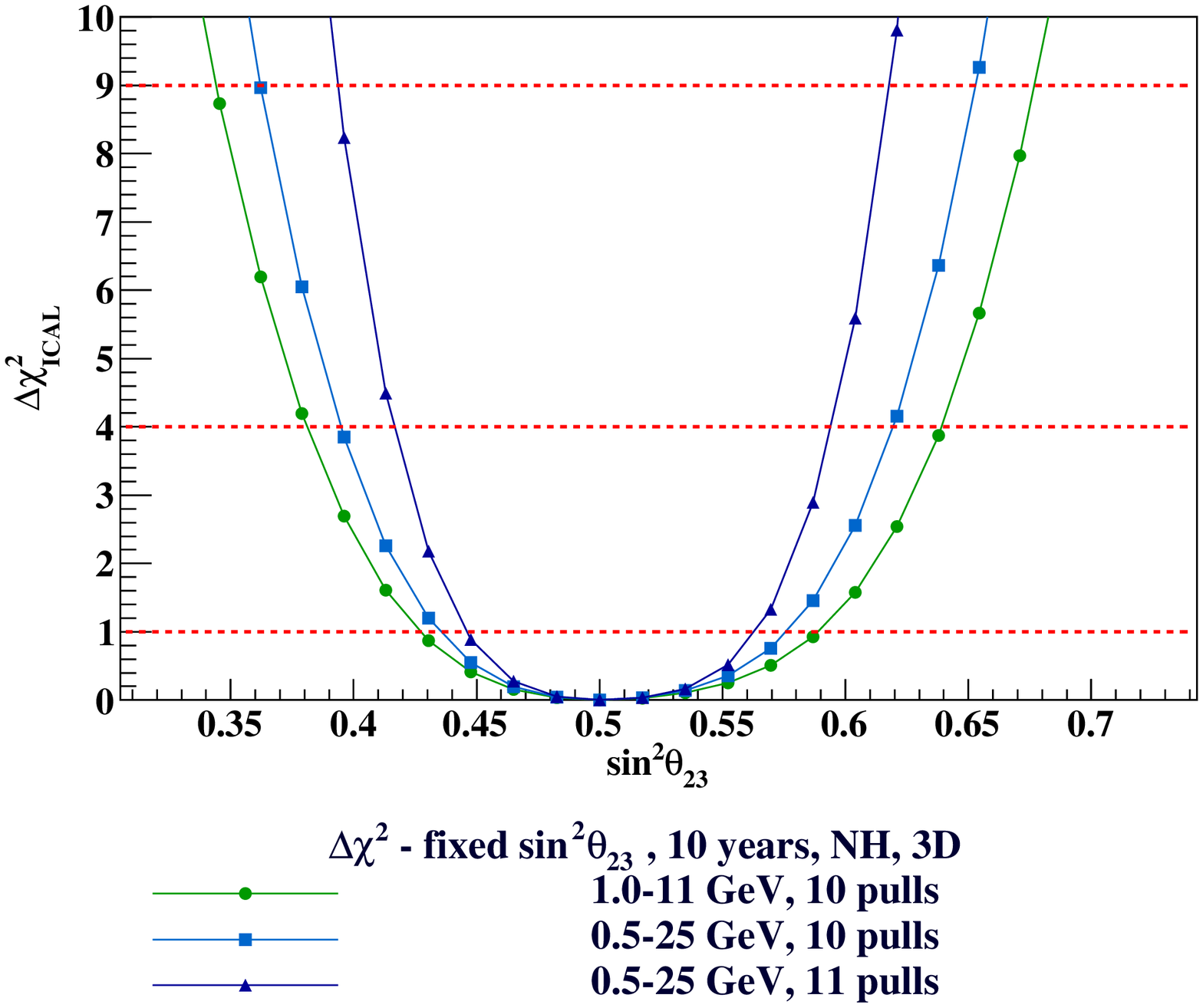}
\caption{$\Delta\chi^2_{\rm ICAL}$ at different values of
$\sin^2\theta_{23}$ with true $\sin^2\theta_{23}=0.5$ and with normal
hierarchy as the true hierarchy. The left panel shows the results for muon
only (2D) analysis and the right panel shows those for the analysis
with hadrons also (3D), for all possible combinations of
energy ranges, pulls and binning schemes.}
\label{pre-parabolas-stt23}
\end{figure}

The relative 1$\sigma$ precision on $\sin^2\theta_{23}$ obtained
from different analyses, with normal hierarchy as the true hierarchy, is
shown in Fig.~\ref{pre-parabolas-stt23} for different cases which are
the combinations of energy ranges, binning schemes and number of pulls.

The other parameters $|\Delta{m^2_{\rm eff}}|$ and $\theta_{13}$
have been marginalised over their 3$\sigma$ ranges as given in
Table~\ref{osc-par-3sig}. Percentage precisions on $\sin^2\theta_{23}$
at 1$\sigma$ obtained with different analyses are shown in
Table~\ref{pre-tab-stt23}.

\begin{table}[htb]
\begin{tabular}{|c|c|c|c|c|c|}
\hline
Binning & $E^{obs}_{\mu}$ (GeV) & No.of pulls & $\sin^2\theta_{23_{min}}$ & $\sin^2\theta_{23_{max}}$ & Precision \\
& & & at 2$\sigma$ & at 2$\sigma$ & at 1$\sigma$ (\%) \\
\hline
&	1--11	&	10	&	0.370	&	0.658	&	14.40 \\
{\color{black}2D (NH)}	&	0.5--25	&	10	&	0.380	&	0.640	&	13.00 \\
{\color{black}$(E^{obs}_\mu,\cos\theta^{obs}_\mu)$}	&	0.5--25	&	11	&	0.412	&	0.599	&	9.35 \\
\hline
&	1--11	&	10	&	0.381	&	0.639	&	12.85	\\
{\color{black}3D (NH)}	&	0.5--25	&	10	&	0.394	&	0.619	&	11.25	\\
{\color{black}$(E^{obs}_\mu,\cos\theta^{obs}_\mu,E'^{obs}_{had})$}	&	0.5--25	&	11	&	0.416	&	0.594	&	8.90	\\
\hline
	&		&		&		&		&		\\
{\color{black}3D (IH)}	&	0.5--25	&	11	&	0.421	&	0.606	&	9.25	\\
{\color{black}$(E^{obs}_\mu,\cos\theta^{obs}_\mu,E'^{obs}_{had})$}  &		&		&		&		&		\\
\hline
\end{tabular}
\caption{Precision of $\sin^2\theta_{23}$ at 1$\sigma$, obtained using Eq.~\ref{sig1-pre}, from different analyses. The maximum and minimum values of 
$\sin^2\theta_{23}$ at 2$\sigma$ in each case are also shown. The true value of $\sin^2\theta_{23}$ is taken to be 0.5 with normal hierarchy as the true 
hierarchy. The last row shows the precision with IH as the true hierarchy.}
\label{pre-tab-stt23}
\end{table}

\paragraph{2D Analysis of $\sin^2\theta_{23}$}: The extension of the $E^{obs}_\mu$ range 
to 0.5--25 GeV improves the precision to 13\% from 14\% obtained in the earlier analysis with 
1--11 GeV \cite{TApre,3dMMD}. A very {\emph {large}} enhancement in precision to 9\%
is obtained when the 11$^{th}$ pull is included. This is a {\emph {very significant observation}}
for all magnetised detectors and the reason for this improvement is discussed in Section~\ref{eff-11-pull}.

\paragraph{3D Analysis of $\sin^2\theta_{23}$}: A similar result of 9\% precision is obtained with 
the 3D analysis. This is better than the earlier result of 13\%
\cite{3dMMD}. The 2D muon-only analysis 
gives a comparable result; this is significant because the muon-only analysis avoids the problem of the 
large uncertainties arising from the mis-identification of hadron hits as muon hits in the detector 
(and vice versa). Of course an improved measurement of hadron energy can further improve this result.

\begin{figure}[htb]
\centering
\includegraphics[width=0.49\textwidth]{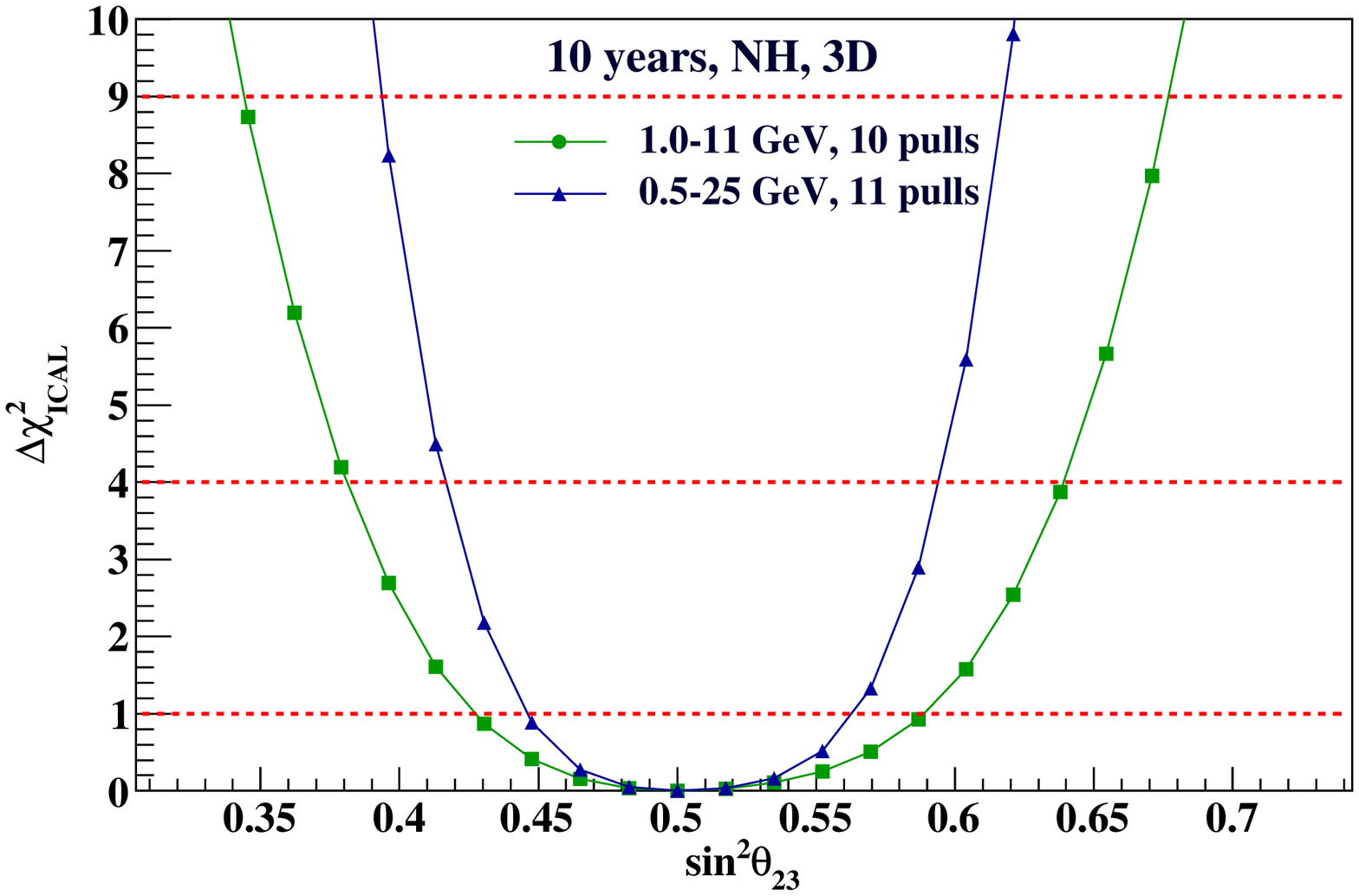}
\includegraphics[width=0.50\textwidth]{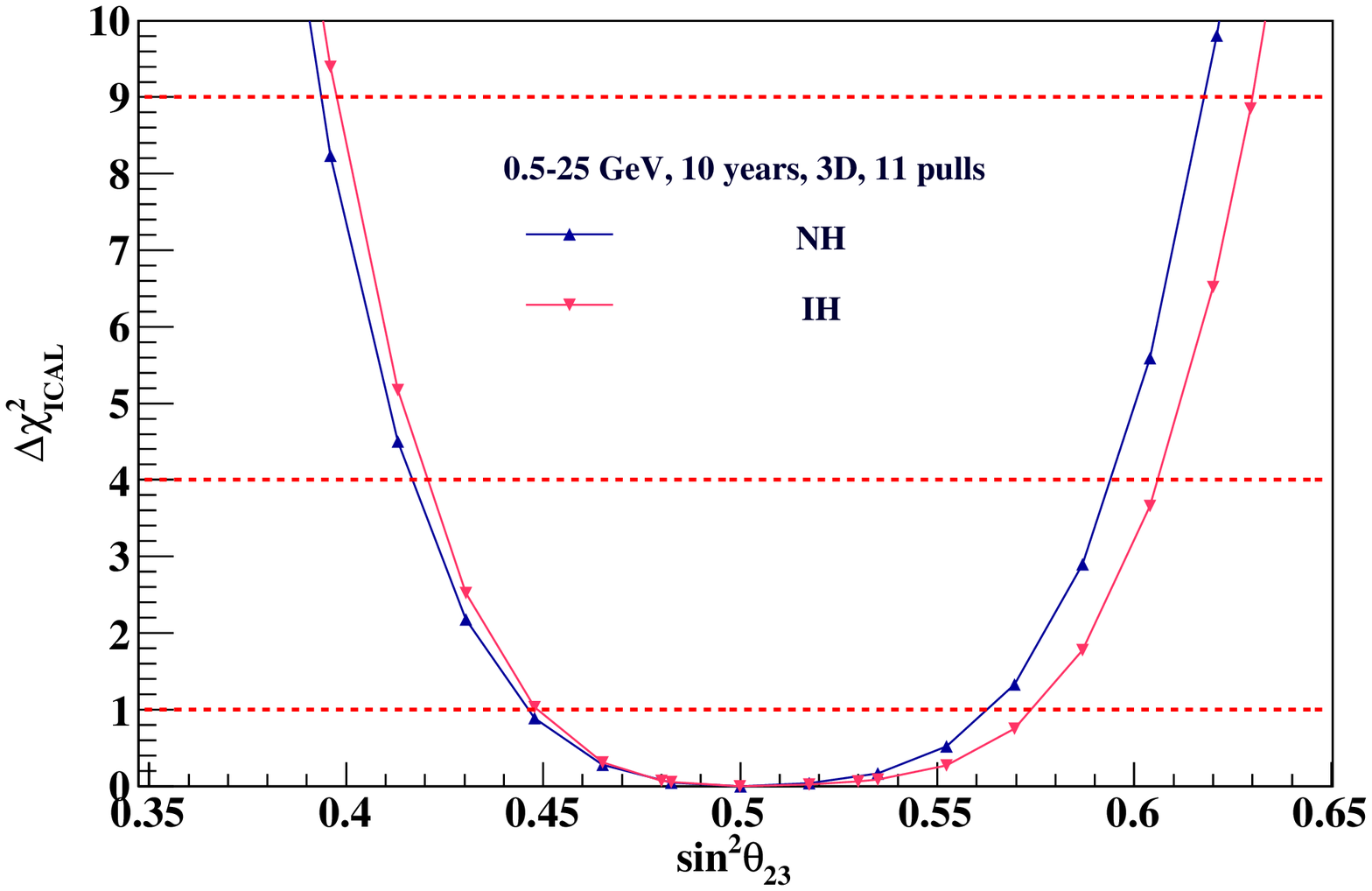}
\caption{Left: The best precision on $\sin^2\theta_{23}$ obtained from the
current in comparison with earlier \cite{3dMMD} 3D simulations analyses
of ICAL, including information from hadrons, assuming normal hierarchy as
the true one. Right: The best precision on $\sin^2\theta_{23}$ obtained
from the current analysis for both hierarchies. For details see text.}
\label{fig:comp-tt23}
\end{figure}

The earlier best result (with 1--11 GeV in muon energy, hadron bins as
listed in Table~\ref{mo-wh-bins}, and including only 10 pulls)
\cite{3dMMD} is shown in comparison with the best results of the present
analysis (using 0.5--25 GeV in muon energy, the same hadron bins, but
with 11 pulls) in Fig.~\ref{fig:comp-tt23} (left) for NH. Indeed,
it can be seen that the earlier best result to $3\sigma$ precision
is worse than that already known from other experiments as listed
in Table~\ref{osc-par-3sig}. The precision obtainable with IH as the 
true hierarchy was also studied with our latest analysis. The relative 
1$\sigma$ precision obtained with IH is 9.25\% which is comparable to 
that obtained from NH (8.9\%), as can be seen in Fig.~\ref{fig:comp-tt23}
(right). This is comparable to or slightly better than the precision on 
$\sin^2\theta_{23}$ obtained by T2K \cite{T2K-latest,T2K-anu,T2K-acc-reac}.
That is, the 11$^{th}$ pull acts as a constraint on the relative $\mu^+$ and 
$\mu^-$ events so that a magnetised atmospheric neutrino detector can achieve 
the precision obtained by an accelerator experiment. We believe that this 
fact has been pointed out for the first time in this paper.

\subsection{Precision on $|\Delta{m^2_{32}}|~(|\Delta{m^2_{31}}|)$}\label{pre-dmeff2}
\begin{figure}[htb]
\centering
\includegraphics[width=0.45\textwidth]{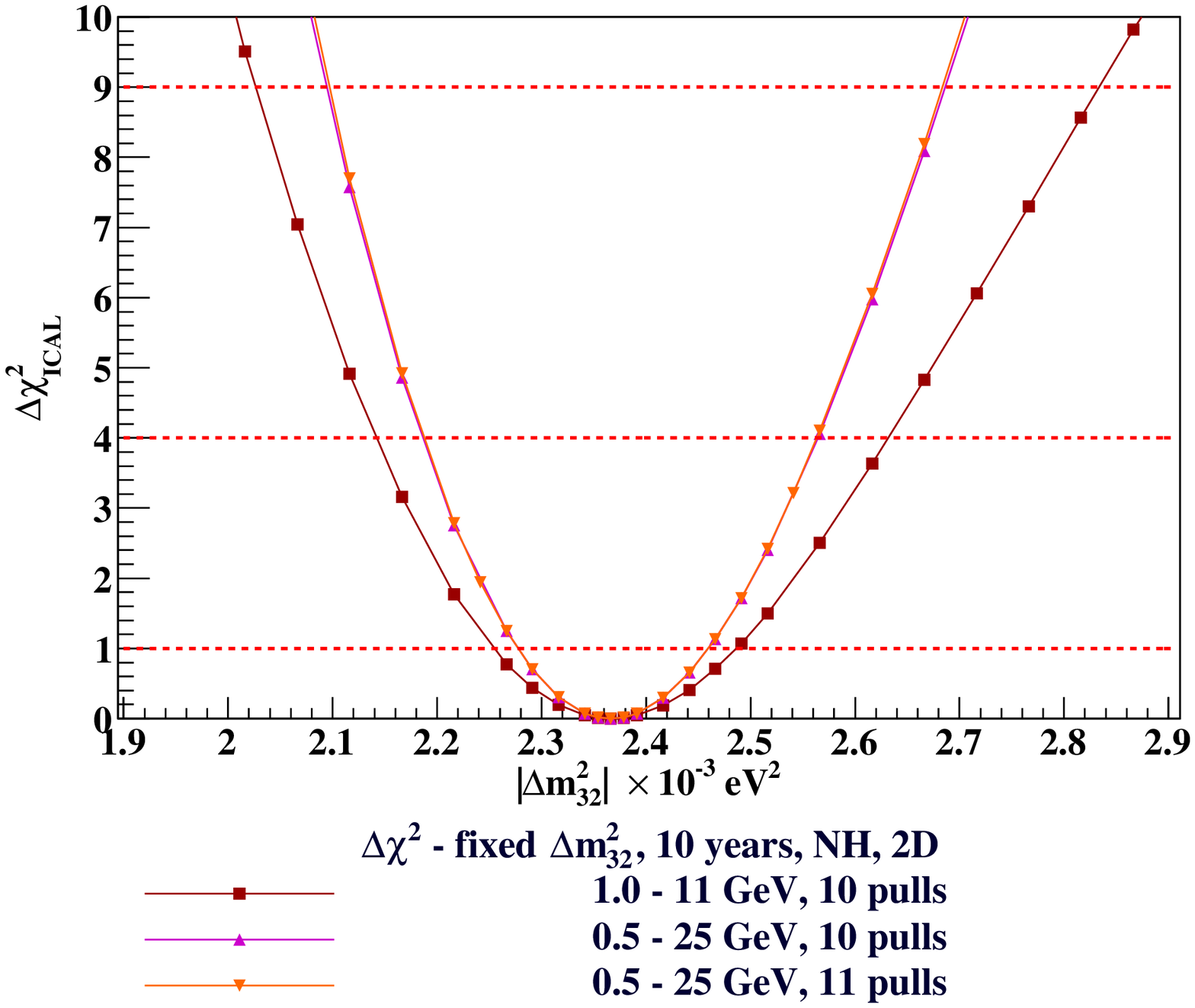}
\hfill
\includegraphics[width=0.45\textwidth]{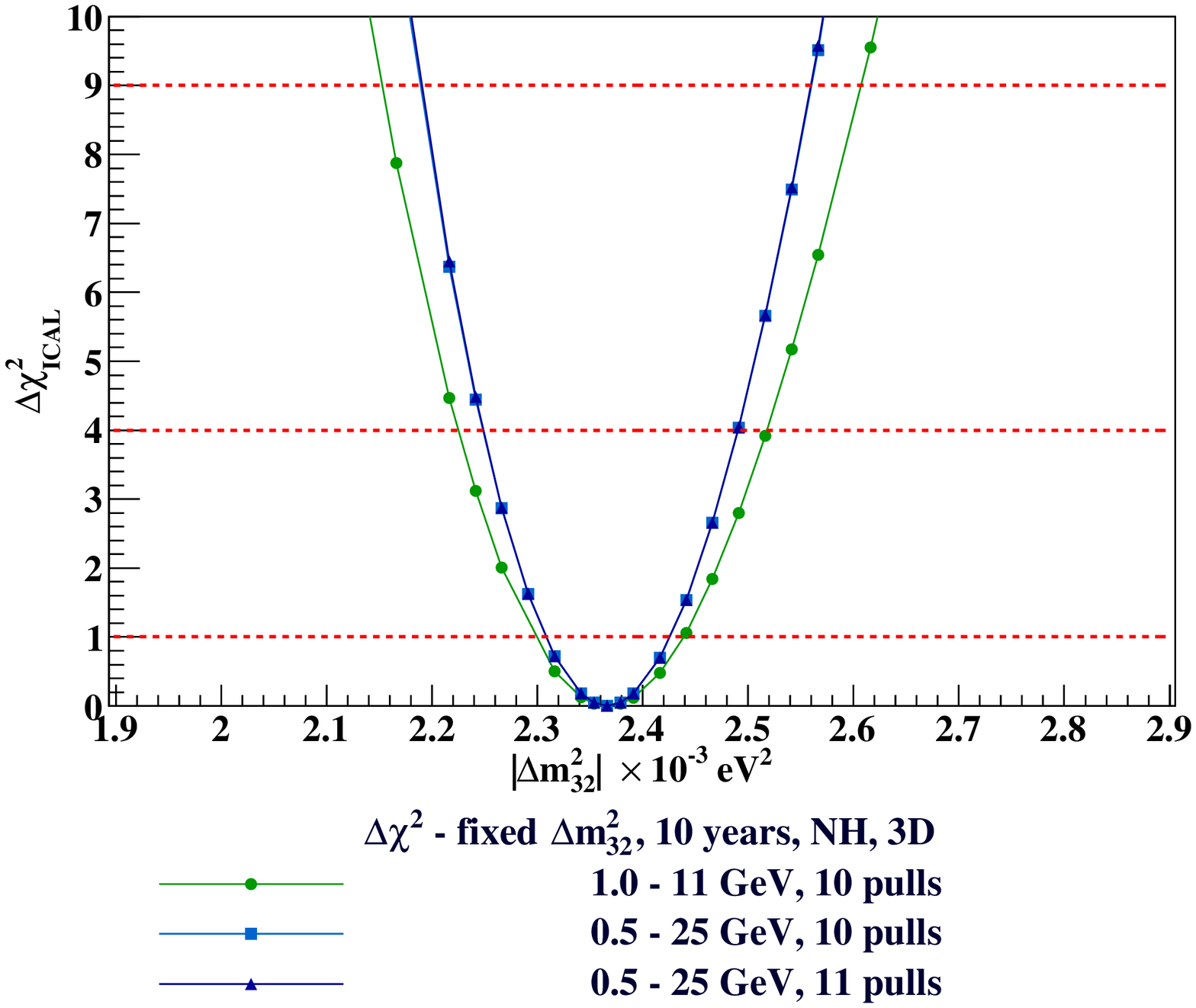}
\caption{Precision on $|\Delta{m_{32}^2}|$ with true
$|\Delta{m_{32}^2}| = 2.366\times10^{-3}~{\rm eV}^2$
($|\Delta{m_{\rm eff}^2}| = 2.4\times10^{-3}~{\rm eV}^2$) with normal
hierarchy as the true hierarchy. The left panel shows the results for
the 2D analyses and the right one shows those for the 3D analyses. The
results for all possible combinations of energy ranges, pulls and
binning schemes are shown.}
\label{pre-parabolas-dmeff2}
\end{figure}

Since ICAL is a magnetised iron calorimeter, it can measure
$|\Delta{m^2_{32}}|$ with very good precision. As in the case of $\sin^2\theta_{23}$, there
are six different analyses which give the results as shown
in Fig.~\ref{pre-parabolas-dmeff2}. The percentage precisions at $1\sigma$
obtained for the magnitude of $\Delta{m^2_{32}}~(\Delta{m^2_{31}})$ are shown in
Table~\ref{pre-tab-dmeff2} when the true hierarchy is normal (inverted).
It can be seen that ICAL will be able to determine 
$|\Delta{m^2_{32}}|~(|\Delta{m^2_{31}}|)$ with a greater precision than
$\sin^2\theta_{23}$, in all energy ranges.

\begin{figure}[htb]
\centering
\includegraphics[width=0.41\textwidth]{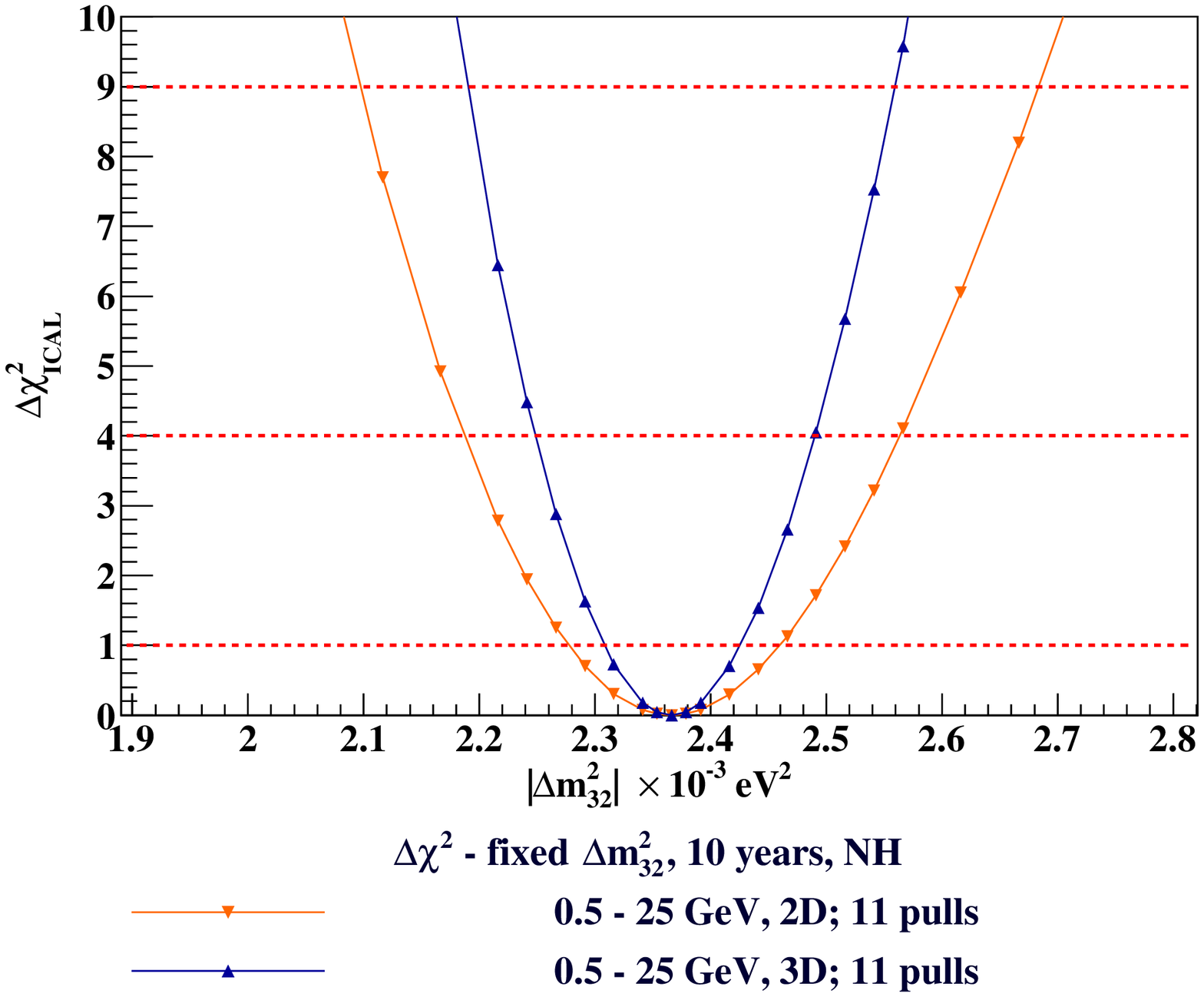}
\includegraphics[width=0.41\textwidth]{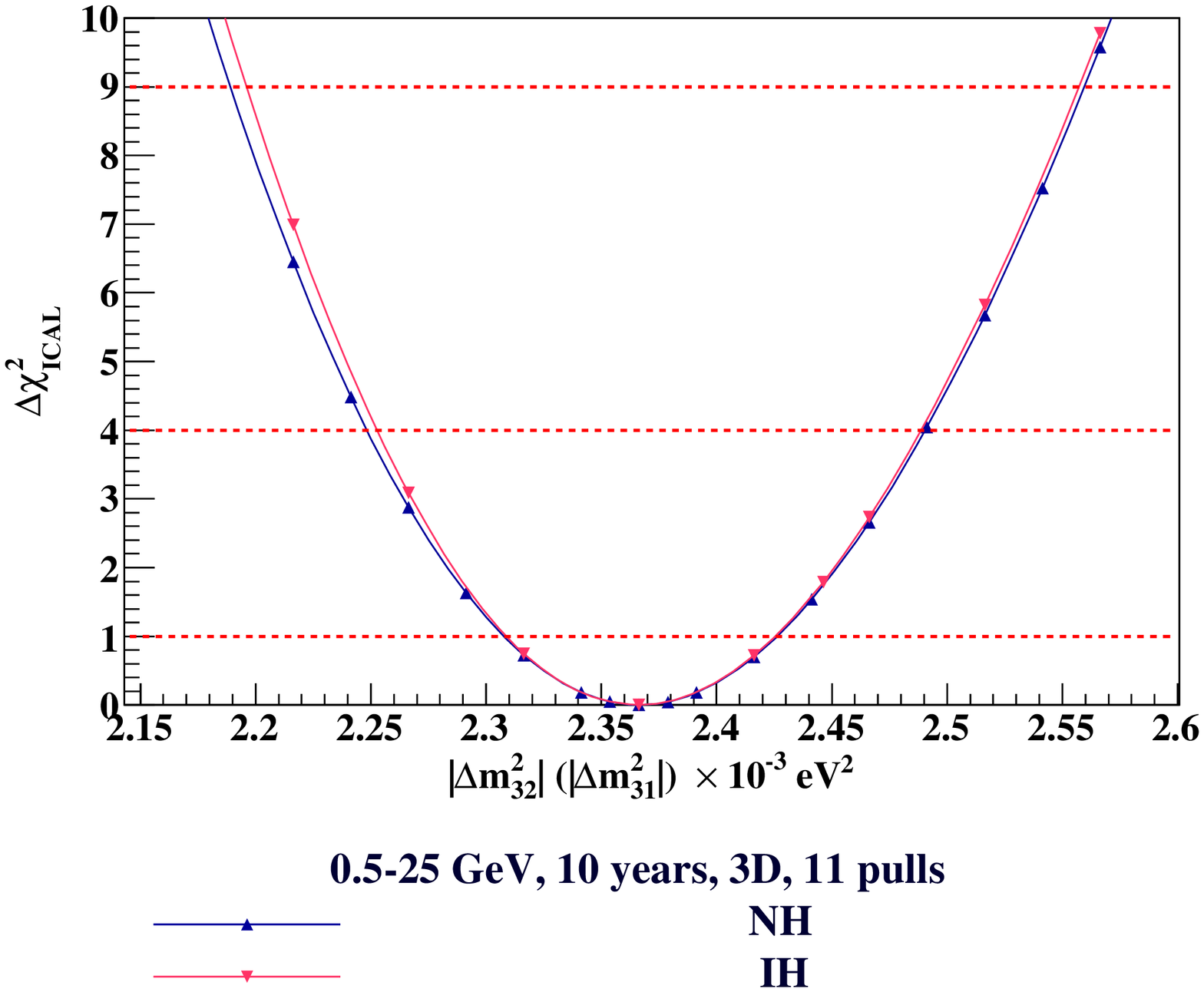}
\caption{Left: Comparison of $\Delta\chi^{2}_{ICAL}$ obtained from 2D and 3D
analyses in the energy range $E^{obs}_\mu$ = 0.5--25 GeV and with 11
pulls with normal hierarchy as the true one. The best result is for the
3D analysis (which is similar with either 10 or 11 pulls. Right: Best
sensitivity to $|\Delta{m^2_{32}}|$ in the current analysis (3D,
11 pulls) for both normal and inverted hierarchies.}
\label{fig:dm2-2d-3d}
\end{figure}

The current best results with 2D and 3D analyses are shown in
Fig.~\ref{fig:dm2-2d-3d} (left). The new 2D and 3D analyses in the range 
$E^{obs}_\mu$ = 0.5--25 GeV constitute an improvement over the 
older 2D and 3D analyses \cite{3dMMD}. However, as the precision on $|\Delta{m^2_{32}}|$
achievable by ICAL is alreay quite good, the improvement does not seem to be as
pronounced as in the case of the case of $\sin^2\theta_{23}$ although the 3D analysis is itself 
an improvement over the current bounds. 

\paragraph{2D Analysis of $|\Delta{m^2_{32}}|$} : The muon-only (2D) analysis with 10 pulls
gives a precision of 4\%, which is a $\sim$ 22\% improvement over the old value. With the additional constraint of 
the 11$^{th}$ pull in the $E^{obs}_\mu$ = 0.5--25 GeV case, the precision achievable is similar, which 
shows that the new pull does not improve the precision further.
This is in contrast to the precision measurement of $\sin^2\theta_{23}$, the precision of which is impacted 
mainly by the constraint on the $\nu_\mu$-$\overline{\nu}_\mu$ flux ratio.The reason for this is discussed 
in Section~\ref{eff-11-pull}.

\paragraph{3D Analysis of $|\Delta{m^2_{32}}|$} : The precision obtained with 3D binning 
and 10 pulls in 0.5--25 GeV improves to 2.5\% from the older value of 3\%, which corresponds 
to a $\sim$ 17\% increase. The addition of the 11$^{th}$ pull again does
not improve the precision further.
Precision measurement with IH as the true hierarchy gives a precision which is comparable
to that obtained for NH. This is shown in the right panel of Fig.~\ref{fig:dm2-2d-3d}, where
the $x$-axis corresponds to $\Delta{m^2_{32}}$ for NH and
$(-\Delta{m^2_{31}})$ for IH.

\begin{table}[h]
\begin{tabular}[hp]{|c|c|c|c|c|c|}
\hline
Binning & $E^{obs}_{\mu}$ (GeV)  & No.of pulls & $|\Delta{m^2_{32(1)}}|_{min}$ & $|\Delta{m^2_{32(1)}}|_{max}$ & Precision \\
& &  & $\times10^{-3}{\rm eV}^2$ at 2$\sigma$ & $\times10^{-3}{\rm eV}^2$ at 2$\sigma$ & at 1$\sigma$ (\%) \\
\hline
&	1--11	&	10	&	2.142	&	2.630	&	5.15 \\
{\color{black}2D}&	0.5--25	&	10	&	2.186	&	2.565	&	4.00 \\
{\color{black}$(E^{obs}_\mu,\cos\theta^{obs}_\mu)$}&	0.5--25	&	11	&	2.188	&	2.563	&	3.96 \\
\hline
&	1--11	&	10	&	2.224	&	2.517	&	3.09 \\
{\color{black}3D}	&	0.5--25	&	10	&	2.248	&	2.491	&	2.57 \\
{\color{black}$(E^{obs}_\mu,\cos\theta^{obs}_\mu,E'^{obs}_{had})$}&	0.5--25	&	11&	2.248	&	2.491	&	2.57 \\
\hline
	&		&		&		&		&		\\
{\color{black}3D (IH)}	&	0.5--25	&	11	&	2.253	&	2.488	&	2.48	\\
{\color{black}$(E^{obs}_\mu,\cos\theta^{obs}_\mu,E'^{obs}_{had})$}  &		&		&		&		&		\\
\hline
\end{tabular}
\caption{Precision of $|\Delta{m^2_{32}}|~(|\Delta{m^2_{31}}|)$ at 1$\sigma$, obtained using
Eq.~\ref{sig1-pre}, from different analyses. The maximum and minimum
values of $|\Delta{m^2_{32}}|~(|\Delta{m^2_{31}}|)$ at 2$\sigma$ in each case are also shown. The
true value of $|\Delta{m^2_{32}}|~(|\Delta{m^2_{31}}|)$ is taken to be 
$|\Delta{m^2_{32}}|~(|\Delta{m^2_{31}}|) = 2.366\times 10^{-3}$ eV$^2$ 
($|\Delta{m^2_{\rm eff}}| = 2.4\times10^{-3}
{\rm eV}^2$) with normal (inverted) hierarchy as the true hierarchy.}
\label{pre-tab-dmeff2}
\end{table}
\subsection{Simultaneous precision on $\sin^{2}\theta_{23}$ and
$|\Delta{m^{2}_{32}}|$}\label{2D-contour}
The results discussed in the previous sections were for fixed values
of either of the oscillation parameters $\sin^2\theta_{23}$ and
$|\Delta{m^2_{32}}|~(|\Delta{m^2_{31}}|)$. We now discuss the parameter space
allowed by our latest analysis. The analysis was done for the 11 pull case with 
$E^{obs}_\mu$ = 0.5--25 GeV and with hadrons, for 500 kton yrs of ICAL exposure. 
Normal hierarchy is assumed to be the true hierarchy. 

\begin{figure}[htp]
\centering
\includegraphics[width=0.45\textwidth]{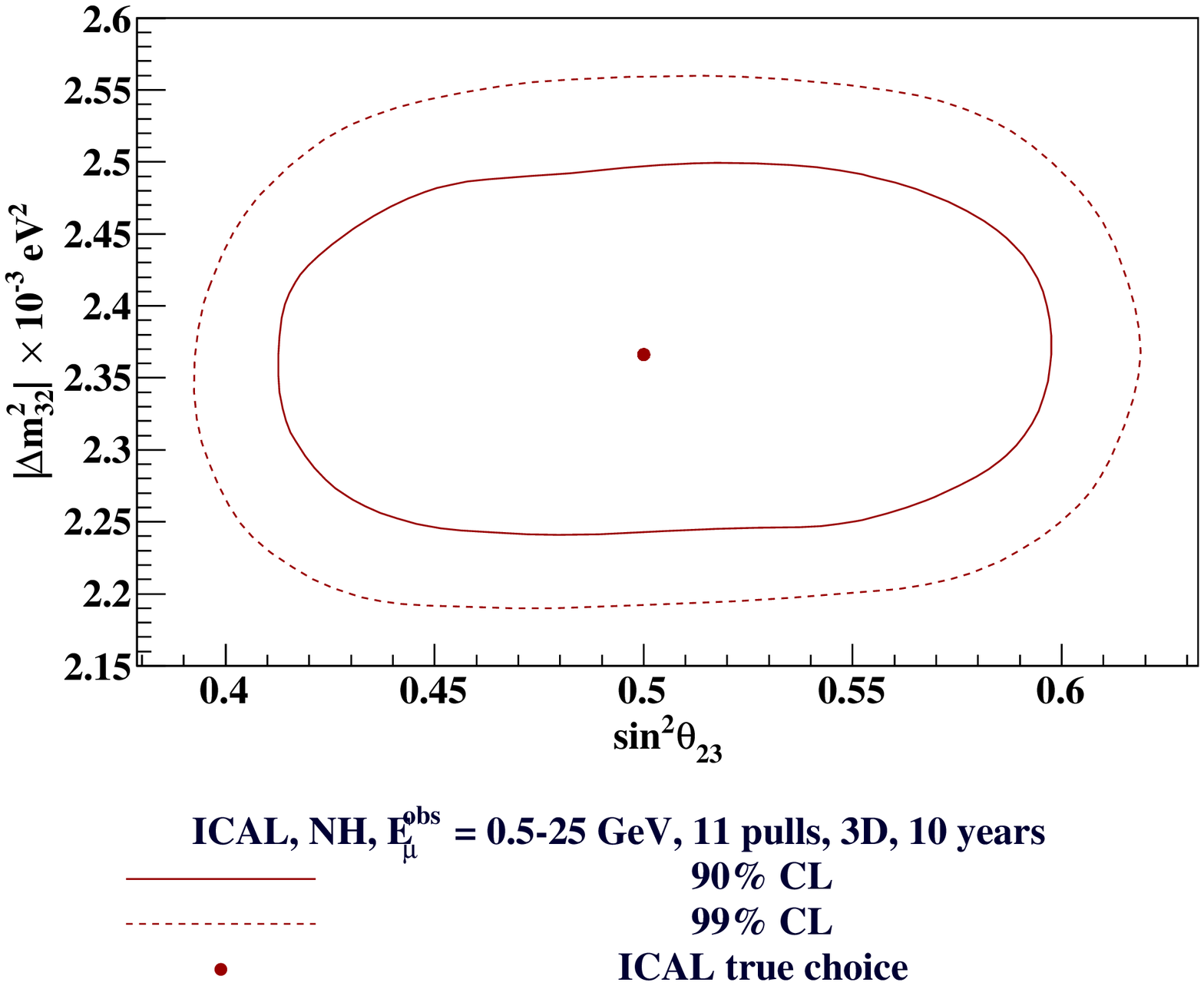}
\includegraphics[width=0.45\textwidth]{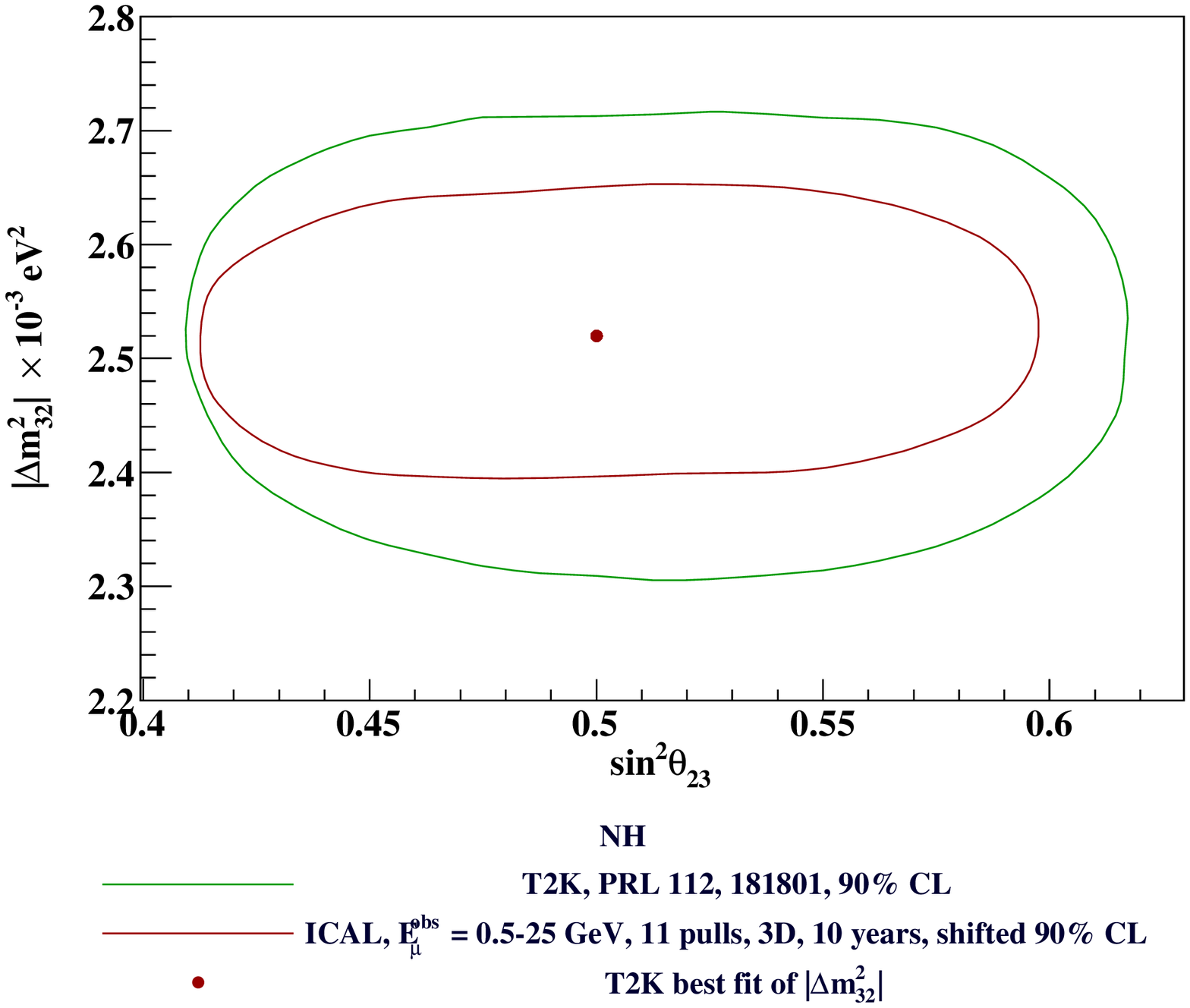}
\caption{(Left) Allowed contours at 90\% and 99\% confidence level for 500 kton yr exposure of ICAL 
in the $\sin^2\theta_{23}-|\Delta{m^2_{32}}|$ plane. The red dot shows the true choice of parameters for the ICAL
analysis, $(\sin^2\theta_{23},|\Delta{m^2_{32}}|)=(0.5, 2.366
\times10^{-3} {\rm eV}^2)$ with NH.
(Right) Comparison of the ICAL projected 10 year sensitivity with the current T2K result at 90\% CL. The ICAL 
contour is shifted to match the best-fit value of $\vert \Delta m^2_{32}
\vert$ from T2K \cite{t2k-2014}.
}
\label{ical-t2k-comp}
\end{figure}

\begin{figure}[htp]
\centering
\includegraphics[width=0.50\textwidth]{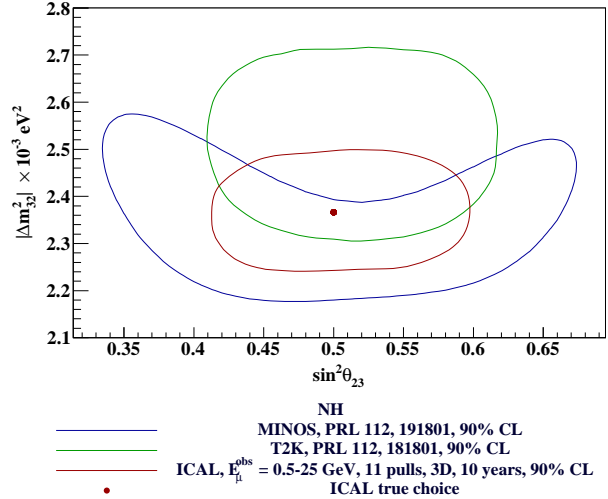}
\caption{The $\Delta\chi^{2}_{ICAL}$ contours at 90\% C.L. 
in the $\sin^2\theta_{23}$--$|\Delta m^2_{32} |$ plane for 500
kton yrs exposure of ICAL (3D analysis with $E^{obs}_\mu$ = 0.5--25 GeV
and the inclusion of the 11th pull). The 90\% confidence contours from
MINOS \cite{minos-numu-nue} and T2K \cite{t2k-2014} are also shown for
comparison. The red dot shows the true choice of parameters for the ICAL
analysis, $(\sin^2\theta_{23},|\Delta{m^2_{32}}|)=(0.5, 2.366 \times
10^{-3} {\rm eV}^2)$.}
\label{cl-pre}
\end{figure}

The 90\% and 99\% confidence contours for 500 kton year exposure of ICAL for NH with the true choices of 
$(\sin^2\theta_{23},|\Delta{m^2_{32}}|) = (0.5, 2.366 \times10^{-3} {\rm eV}^2))$
are shown in the left panel of Fig.~\ref{ical-t2k-comp}. Similar results hold true for 
inverted hierarchy as the true hierarchy. It can be seen that the extension of the energy range for analysis and
constraining the $\nu_\mu-\overline{\nu}_\mu$ flux ratio in ICAL result
in an improved 
sensitivity to the precision on both parameters. The projected ICAL precision on 
$\sin^2\theta_{23}$ is better than or comparable to the current T2K precision as can be seen from the 
right panel of Fig.~\ref{ical-t2k-comp}, where the 90\% CL contour is compared to the results from 
T2K \cite{t2k-2014}.
A comparison of these projected results for
ICAL (NH) with the current results from MINOS \cite{minos-numu-nue} and T2K
\cite{t2k-2014} are shown in Fig.~\ref{cl-pre}. It must be remembered,
though, that 
these experiments are already taking data while ICAL is yet to be constructed!

\subsection{Sensitivity to neutrino mass ordering}\label{hie}
\begin{figure}[htb]
\centering
\includegraphics[height=0.42\textwidth]{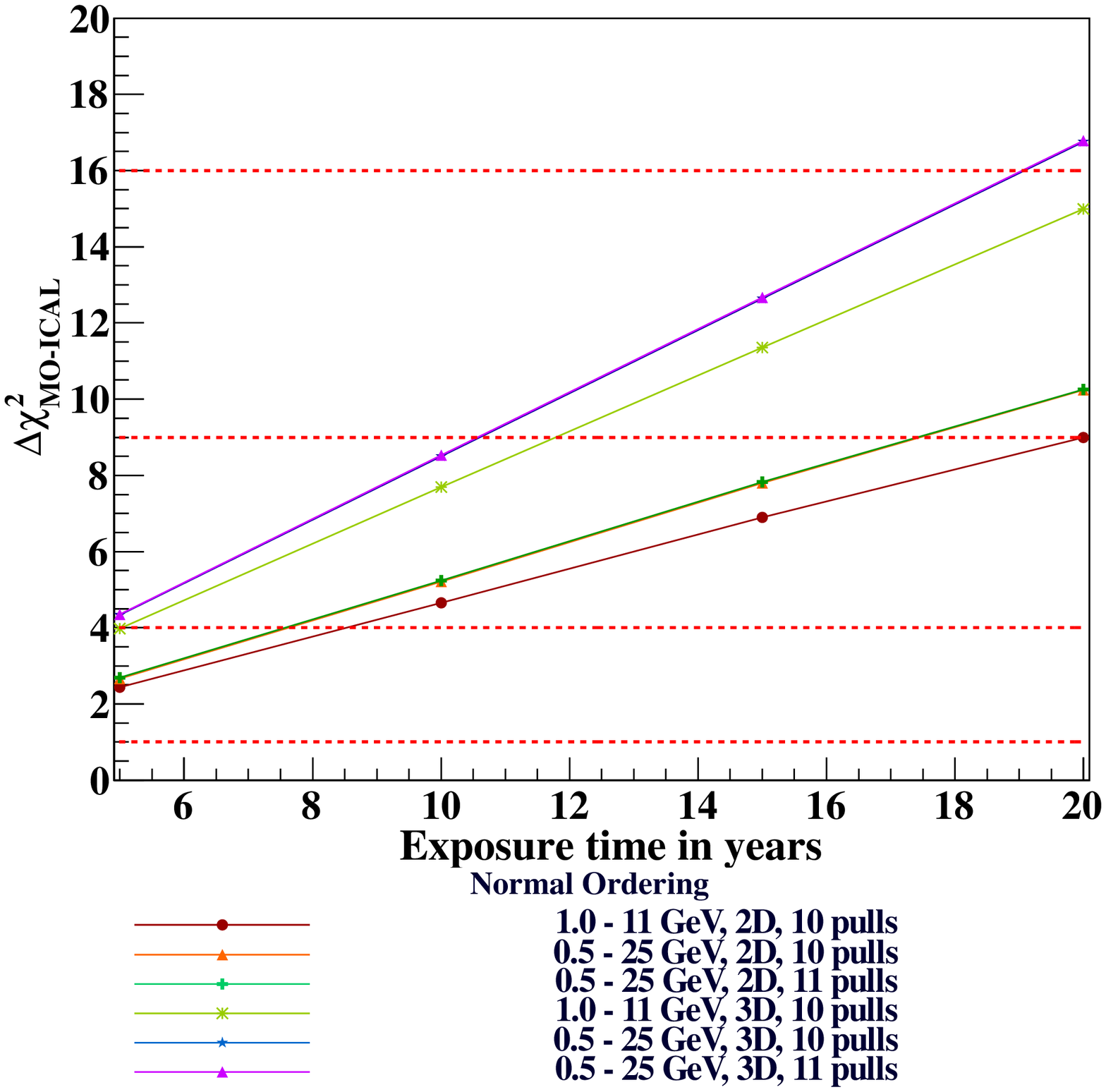}
\hfill
\includegraphics[height=0.42\textwidth]{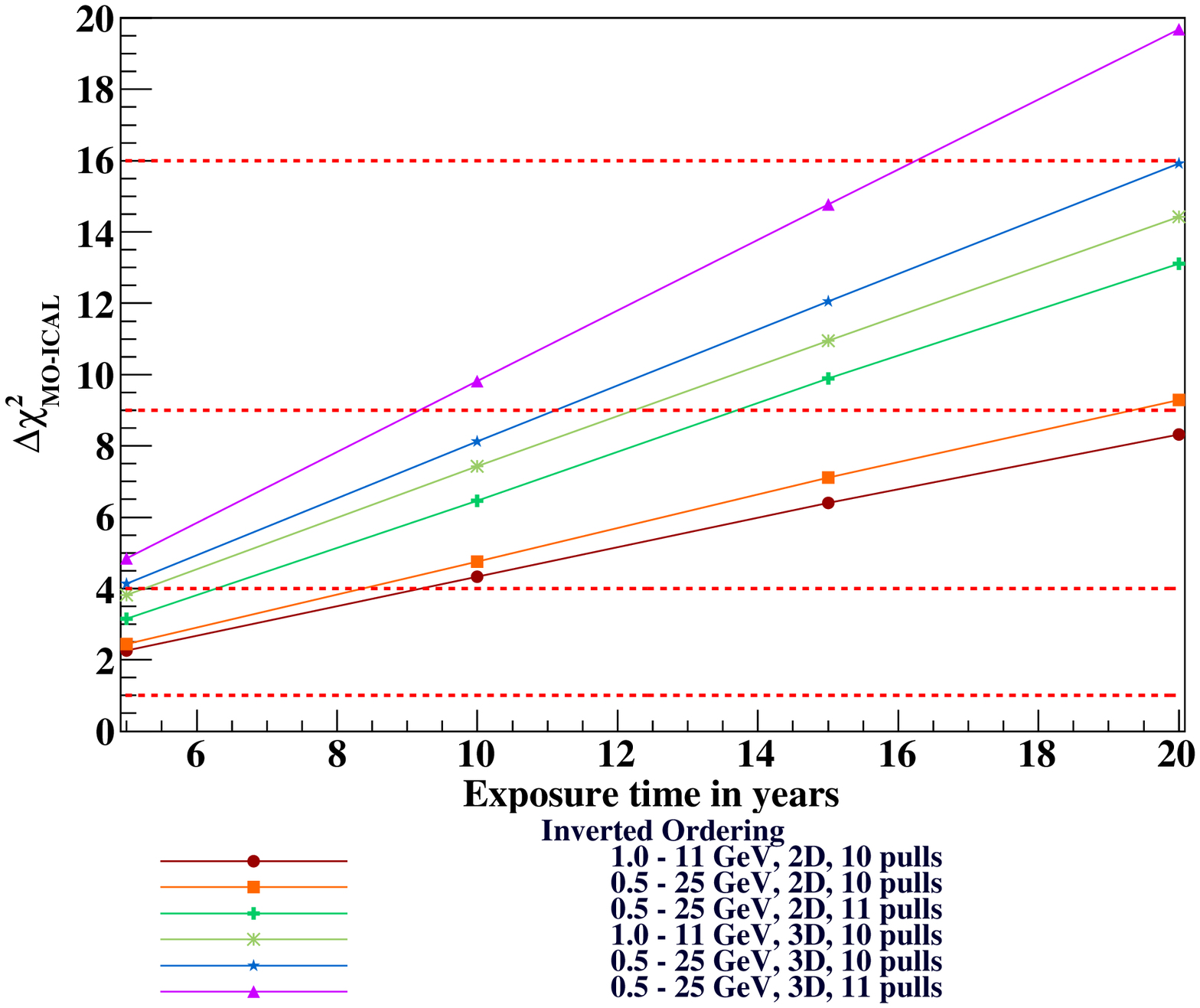}
\caption{$\Delta\chi^2_{MO\hbox{-}ICAL}$ as a function of exposure time
in years. The extension of observed muon energy range
to 0.5--25 GeV increases the sensitivities in the case of both 2D and 3D.
The extra pull improves the hierarchy sensitivity when the true ordering
is inverted.} 
\label{hie-plots}
\end{figure}

ICAL with its magnetisability is an exclusive mass hierarchy machine. 
Most importantly, its ability to discriminate the normal and inverted 
mass ordering is {\em independent} of the CP phase \cite{WP}.
The ability of ICAL to distinguish the correct mass ordering in the 2--3 sector is
given by :
\begin{equation}
\Delta\chi^2_{MO\hbox{-}ICAL} =
\chi^2_{false\hbox{-}MO}-\chi^2_{true\hbox{-}MO}~,
\label{chisq-mh}
\end{equation}
where $\chi^2_{false\hbox{-}MO}$ is the minimum $\chi^2$ calculated
using the false ordering, while allowing $\theta_{23}$, $\theta_{13}$
and $\vert \Delta m^2_{\rm eff} \vert$ to vary over the $3\sigma$
range given in Table~\ref{osc-par-3sig} and $\chi^2_{true-MO}$ is
the minimum $\chi^2$ assuming the true mass ordering. The plot of
$\Delta\chi^2_{MO\hbox{-}ICAL}$ as a function of the number of years of
exposure is shown in Fig.~\ref{hie-plots} both when the true ordering
is taken to be normal (NO) as well as inverted (IO).

\noindent {\bf Normal Ordering}: The muon-only analysis (2D) with 10 pulls only gives
a $\Delta\chi^2_{MO\hbox{-}ICAL}$ of $\sim$ 5.2 for 10 years of ICAL
exposure, better than the earlier result of $\sim$ 4.6, a 13\% increase in the hierarchy sensitivity.
Again the addition of the 11th pull does not improve $\Delta\chi^2_{MO\hbox{-}ICAL}$ in both energy
ranges in the 2D analysis.

The addition of hadron energy as the third observable increases
the $\Delta\chi^2_{MO\hbox{-}ICAL}$ to $\sim 8.5$, for an exposure 
time of 10 years, an improvement over the earlier value of 7.7.
Again the addition of the 11th pull has no effect on the hierarchy sensitivity
when normal ordering is taken to be the true ordering.

\noindent {\bf Inverted ordering}: While trends are similar to that with
normal order as the true order, the inclusion of the 11th pull has
{\emph {significant impact}} on mass hierarchy sensitivity when the true ordering 
is {\emph {inverted}}. In fact, the best sensitivity ($E^{obs}_\mu$ = 0.5--25 GeV,
with 11 pulls and 3D binning) is better with the inverted than with the
normal hierarchy (by $\sim$ 16\%). The reason for this effect with the 11$^{th}$ pull is
discussed in the next section.

It should be noted that the values of $\Delta\chi^2_{MO\hbox{-}ICAL}$
are lower than those reported in Ref.~\cite{3dMMD}, for the same exposure
time. This is due to the fact that the earlier analysis used the input
value $\theta_{13} \sim 9.217^\circ$ while our analysis uses the
current\footnote{The best fit at the time when this analysis was begun.} 
best fit value \cite{pdg-2014} $\theta_{13}\sim 8.729^\circ$. Given that 
the best fit of $\theta_{13}$ has reduced further to $\sim8.5^\circ$ \cite{nufitpage,gonzalezgarcia-2014}
it is even more important to perform the analysis in as wide a kinematic range (energy and direction) 
as possible to get the best possible hierarchy discrimination.

\subsubsection{Dependence on $\theta_{23}$}\label{t23}
\begin{figure}[htp]
\centering
\includegraphics[width=0.47\textwidth]{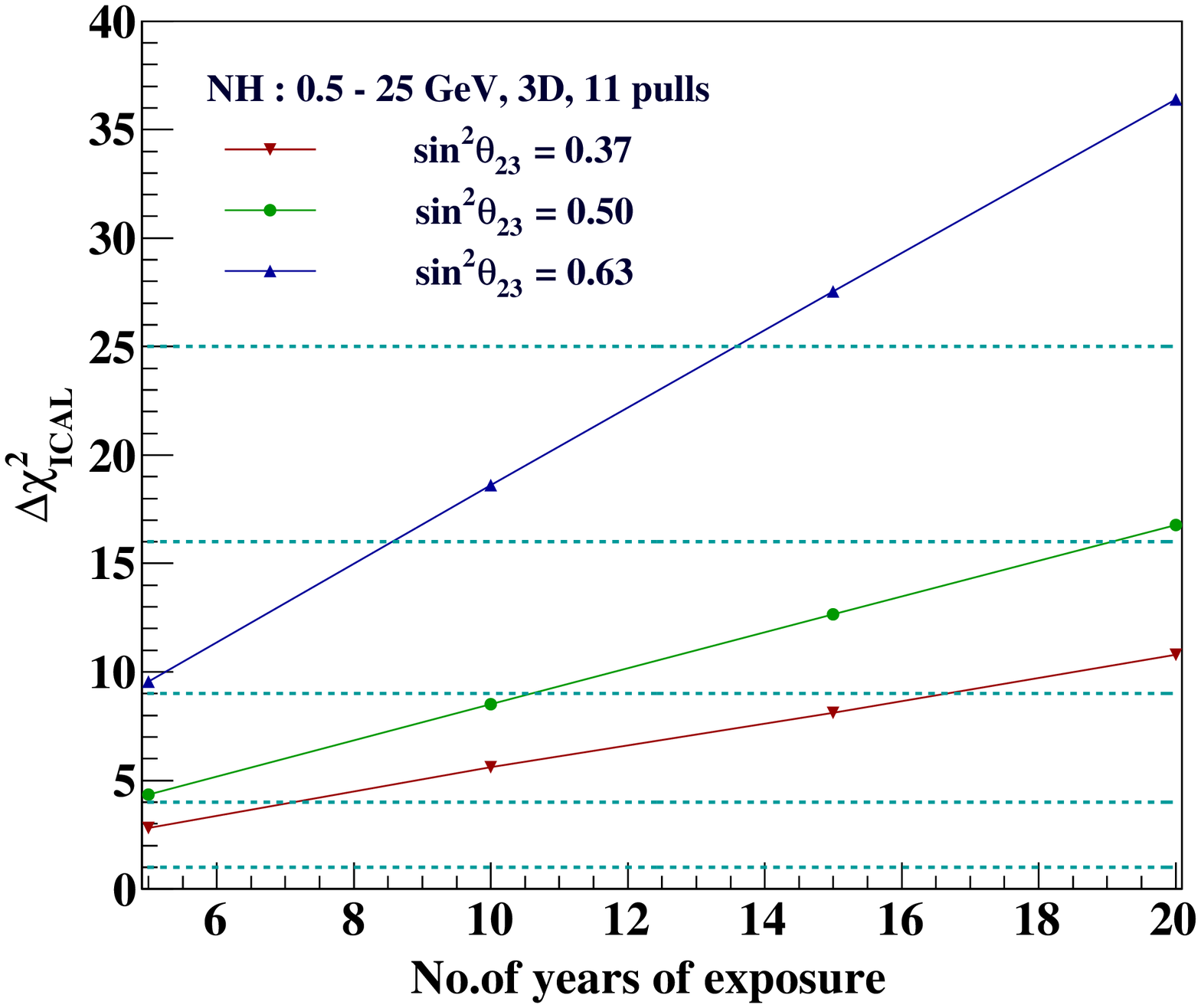}
\hfill
\includegraphics[width=0.47\textwidth]{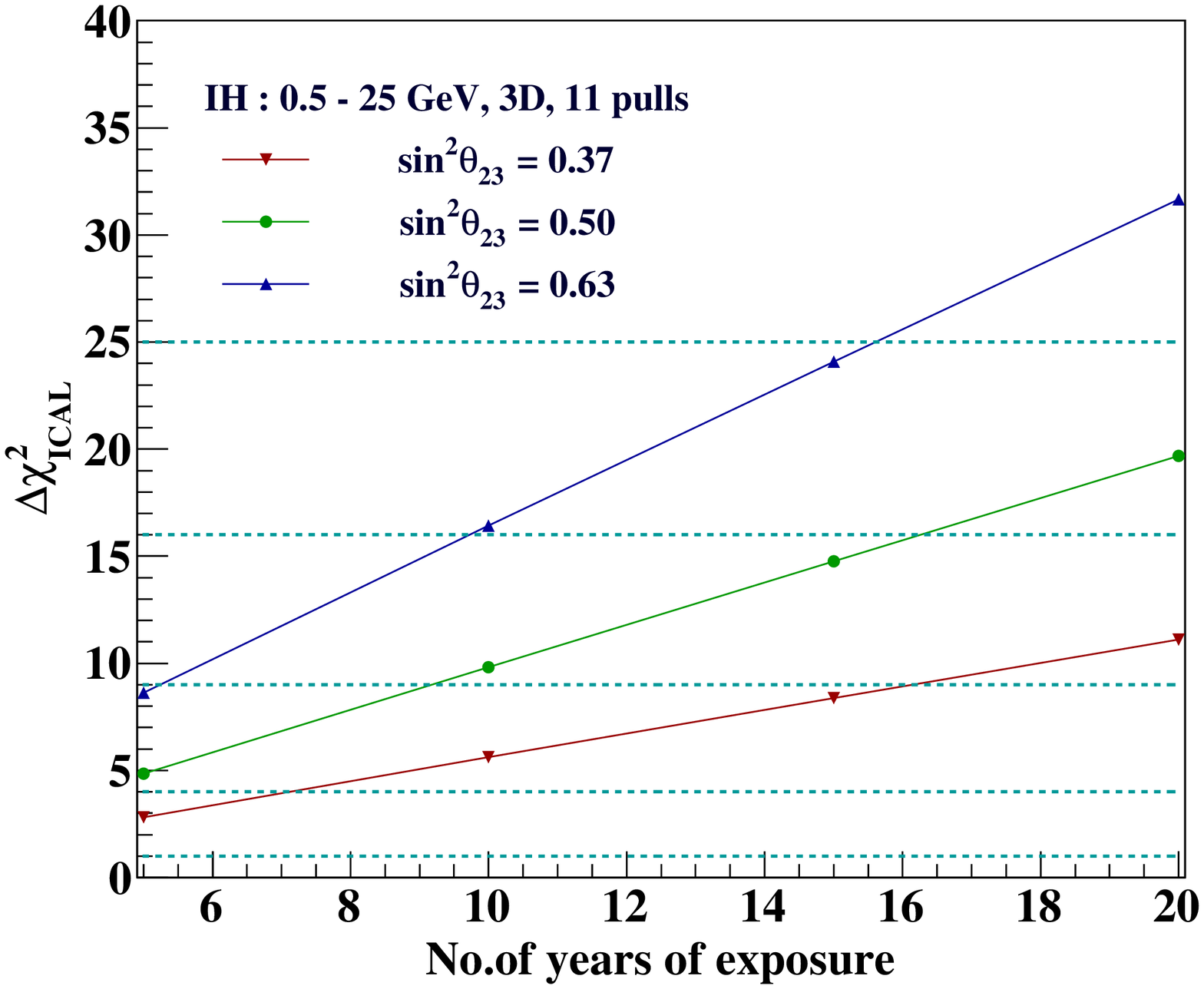}
\caption{The best case values of $\Delta\chi^2_{MO\hbox{-}ICAL}$ as a
function of exposure time in years. It can be seen that the sensitivity
to mass hierarchy increases linearly with the true value of
$\theta_{23}$ when it is varied over the $3\sigma$ range of the parameter.}
\label{fig:hie-23}
\end{figure}

The sensitivity to mass ordering is also known to depend on the true value of $\theta_{23}$; it is higher 
for larger $\theta_{23}$ as can be seen from Fig.~\ref{fig:hie-23}. While ICAL has better sensitivity to 
the inverted ordering when $\theta_{23}$ is in the first octant, the reverse is true when it is in the second 
octant. This is due to the dominance of the
$P_{\mu\mu}$ term in Eq.~\ref{toteve} arising from the survived $\nu_\mu$s
compared to the $P_{e\mu}$ term from oscillated $\nu_e$s, and the nature
of its dependence on $\theta_{23}$. In fact, the results would also be
vastly improved by ``switching off'' the $P_{e\mu}$ term since their
dependence on the oscillation parameters (especially $\theta_{23}$) are
practically the opposite of each other \cite{imsc-par-paper}. This is not
possible with atmospheric neutrinos where Nature provides both flavours,
but is possible with neutrino beams. In the latter case, however, the fact
that there is a single base-line which results in a significant dependence on
and correlation with the CP violating phase $\delta_{CP}$ and complicates
the analysis, as with MINOS \cite{minos-numu-nue}, T2K \cite{T2K-latest},
LBNE \cite{lbne,lbno} or NO$\nu$A \cite{TAhie,NOvA}.

To summarise, the sensitivity to the mass ordering in the 2--3 sector
improves with the addition of higher energy bins in the analysis,
while constraining the $\nu_\mu~-~\overline{\nu}_\mu$ flux ratio
improves the sensitivity only when the true ordering is inverted.
The ICAL's ability to determine this mass ordering is significant
owing to its magnetisability and its 50 kton mass. Improvement
in energy resolutions will further improve the detector's sensitivity to
this parameter. Also, in ICAL, the mass ordering can be determined
independent of the CP violating phase $\delta_{CP}$ because of the range
of
baselines involved in atmospheric neutrinos \cite{WP}; this is in
contrast to beam/short-base-line experiments where there is a non-trivial
sensitivity to the 2--3 mass ordering depending on the true value of the
CP phase. Hence ICAL will be important in the determination of this mass
ordering, and any amount of improvement in determining this parameter is noteworthy.

\section{Impact of the 11$^{\rm{th}}$ pull on determination of the oscillation
parameters}\label{eff-11-pull}

The 11$^{\rm th}$ pull accounts for the fact that the {\em ratios} of the
$\nu_\mu$ and $\overline{\nu}_\mu$ fluxes are better known (to within
5\%) than the absolute fluxes themselves \cite{honda-paper}. This is
implemented by using a pull $\pi_6 = 0.025$, which contributes {\em with
the opposite sign} for neutrino and anti-neutrino events, as can be seen 
from Eq.~\ref{pi6-xi6}.

\begin{figure}[tbp]
\centering
\includegraphics[width=0.70\textwidth]{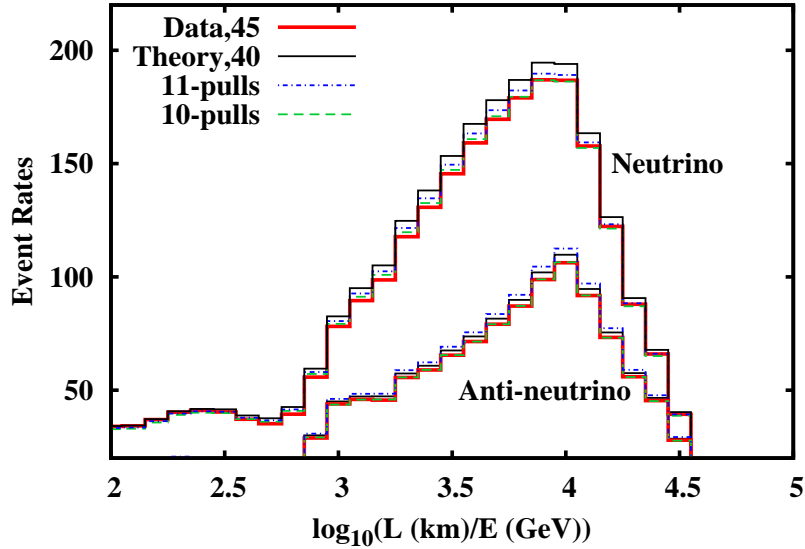}
\caption{The 10 year rates from CC $\nu_\mu$ and $\overline{\nu}_\mu$
events shown as a function of $\log_{10}(L(\hbox{km})/E(\hbox{GeV}))$
of the final state muon. The red (thick solid) lines correspond
to the ``data'' and the black (thin solid) ones to the ``theory"
with $\theta_{23} = 40^\circ$ (and other parameters held fixed). The
green (dashed) lines (practically overlapping the data) and the blue
(dot-dashed) lines mimic the effect of 10th pull and inclusion of the
11th pull respectively.}
\label{fig:pull}
\end{figure}

It is seen that the inclusion of the 11th pull is most visible in the
determination of $\theta_{23}$ which becomes more constrained when this
pull is included. One way to understand this is to re-bin the events in a
single variable $L/E$ (of the final state muon) and consider the effect of 
just this pull. Fig.~\ref{fig:pull} shows the effect of
$\theta_{23}$ on both the neutrino and anti-neutrino events. The thick
solid line is the ``data'' corresponding to $\theta_{23} = 45^\circ$ while
the thin solid ``theory" line corresponds to a fit with $\theta_{23} =
40^\circ$ and without any pull. In both cases, reducing $\theta_{23}$
from the true value increases the event rate {\em in every bin} (the
opposite will hold with the inverted hierarchy; here the normal hierarchy
is shown). Note that the down-going events are not shown here.

The dot-dashed and dashed curves corresponding to the labels 11- and
10-pulls show the effect of changing the (single) normalisation of the
theory with and without the 11$^{\rm th}$ pull. The overall normalisation
of the events in the 10-pull case can be {\em independently} varied
for neutrinos and anti-neutrinos (decreased by 4\% and 3\% in figure)
to improve the agreement of the $40^\circ$ theory line to the data,
resulting in smaller overall $\chi^2$ in this fit. On the other hand,
a 2.5\% decrease in the normalisation of neutrino-induced events in the
11-pull case is accompanied by a 2.5\% {\em increase} in the anti-neutrino
case, so that the agreement with the neutrino data becomes better, but
that with the anti-neutrino data becomes worse. (Of course, it can be
applied vice versa, but the smaller $\chi^2$ is obtained with this choice
since there are about twice as many neutrino events as anti-neutrino
ones due to the smaller cross section of the latter.) Hence it is
not possible to improve the agreement of the $40^\circ$ theory with
data by tuning the normalisation in the analysis when including this pull;
this results in a larger $\chi^2$ compared to the 10-pull case where
the normalisations of neutrino- and anti-neutrino-induced events can
be independently varied. This gives rise to the tighter constraints on
$\theta_{23}$ when the 11th pull is added. We note that only a detector
like ICAL that is capable of charge identification can successfully
implement this pull as a constraint.

It can also be seen from Fig.~\ref{fig:pull} that there is greater
sensitivity to $\theta_{23}$ in the neutrino rather than in the
anti-neutrino sector. The reverse is true with inverted mass ordering.
In the determination of the sensitivity to the mass ordering, the
minimum $\chi^2$ with the false ordering is found. That means, when the
true ordering is inverted, the ``theory'' is obtained using the normal
ordering, where there is greater sensitivity to $\theta_{23}$ as just
mentioned. When the 11th pull is included, therefore, the discrepancy
between theory and data cannot be achieved by changing $\theta_{23}$ and
hence there is more sensitivity to determination of the mass ordering
when the true ordering is inverted, provided $\theta_{23}$ is in
the first octant. The reverse is true when $\theta_{23}$ is in the
second octant, as can be seen from Fig.~\ref{fig:hie-23}.

In addition, it can be seen from Fig.~\ref{fig:pull} that
there is sensitivity to oscillation parameters near and beyond
$\log_{10}(L ({\rm km})/E ({\rm GeV})) \sim 4$. This is precisely the
region that is included when the range of $E^{obs}_\mu$ is extended from
1 GeV down to 0.5 GeV. Similarly, although smaller, sensitivity to the
parameters is also seen for $\log_{10}(L ({\rm km})/E ({\rm GeV}))
\sim 2$--3, which corresponds to the extension in the higher energy end
from 11 to 25 GeV.

\section{Summary and Discussion}\label{summary}
This paper contains a simulation study of the physics potential of
the proposed 50 kton magnetised Iron Calorimeter detector (ICAL) at INO which
aims to probe neutrino oscillation parameters by observing atmospheric
neutrino oscillations and studying their Earth matter effects as they
propagate through the Earth. This will be done by detecting (mainly)
the charged current interactions of $\nu_\mu$ and $\overline{\nu}_\mu$ in
the detector by means of the final state muons. The detector, which
is optimised for the detection of muons in the GeV energy range, will
have a magnetic field which will enable the distinction of $\nu_\mu$
and $\overline{\nu}_\mu$ events by identifying the charge of the muon in
the final state, thus making ICAL an excellent detector to determine
the neutrino mass ordering. Not only this, the magnetic field helps to
improve the precision measurement of the mixing angle $\theta_{23}$ (and
$|\Delta{m^2_{32}}|$ and the mass ordering as well).

The main themes of our study were the effects of extending the observed
muon energy range to
0.5--25 GeV (from 1--11 GeV used in earlier studies) and that of a constraint on the $\nu_\mu$--$\overline{\nu}_\mu$ flux ratio on the 
sensitivity of a 50 kton ICAL to neutrino oscillation parameters in the 2--3 sector.
The second---and the aspect which is found to have the biggest impact so
far on oscillation sensitivity---arises from two facts; one that ICAL detects atmospheric neutrinos and the other that this massive
detector will be magnetised.

In particular, we show that the relatively small uncertainties on the
atmospheric neutrino-antineutrino flux {\em ratios} act as a constraint
in analyses where the neutrino and antineutrino events can be
separated. This is true in a magnetised detector such as ICAL where the
magnetic field distinguishes muons and anti-muons produced in charged
current interactions of neutrinos and antineutrinos respectively with
the detector material. The presence of the magnetic field
enables ICAL to distinguish between neutrino and anti-neutrino events;
hence, inclusion of the uncertainty on the
$\nu_\mu/\overline{\nu}_{\mu}$ flux ratio as an additional pull
translates to a constraint on this ratio which in turn significantly
improves the precision with which $\sin^2\theta_{23}$ can be determined.
Such a constraint is applicable for all magnetised detectors which have
good charge identification capabilities and the
impact of this constraint has been shown for the first time in this
paper. As a consequence, our simulation studies show that, with 10
years of data taking, ICAL will not only be able to determine $\vert \Delta
m^2_{32} \vert$ with good precision (as expected) but can also pin down
$\sin^2\theta_{23}$ to a precision better than the current limits set by
T2K.

The precision that can be achieved on these parameters in about 10
years' running is about 2.5\%
and 9\% for $\vert \Delta m^2_{32}\vert$ and $\sin^2\theta_{23}$
respectively; see details in Tables~\ref{pre-tab-stt23} and \ref{pre-tab-dmeff2}.

The studies presented here assume that there is perfect separation between
different types of charged current (CC) and neutral current (NC) events. The 
event separation efficiency will affect the results of the analysis since they 
will determine the actual number of events in each bin apart from the contamination 
from NC oscillation-independent events. However the inclusion of this consideration 
is beyond the scope of this paper. Preliminary studies \cite{LSM-thesis} show that 
certain selection criteria can be applied so that an event sample which comprises more than
95\% CC muon events can be obtained. The criterion results in cutting out
events with $E_\nu \lesssim 0.5$ GeV; this has determined largely the range
(lower limit) of muon energy analysed in this paper. Also, improvements in
the reconstruction efficiencies and resolutions (that have been studied
with GEANT-4 simulations of the detector) as well as possible changes in
detector geometry can all alter the results of this analysis.  Note that
the current studies were all done with the atmospheric neutrino fluxes
computed at the Super Kamiokande site \cite{honda-paper}. The fluxes at
Theni where ICAL is proposed to be built are slightly different and are smaller
at energies less than 10 GeV \cite{honda-paper,honda-paper-1}; this
will also impact the analysis. 

Even within these limitations,
it appears that the physics results of ICAL will have the capability to
impact global fits to neutrino data and thus any new analysis will open
a window to understanding the neutrino oscillation parameters better on
the whole and the atmospheric neutrino fluxes themselves, in particular.

\appendix
\section{The effect of fluctuations}\label{fluct}
\begin{figure}[thp]
\centering 
\includegraphics[width=0.45\textwidth]{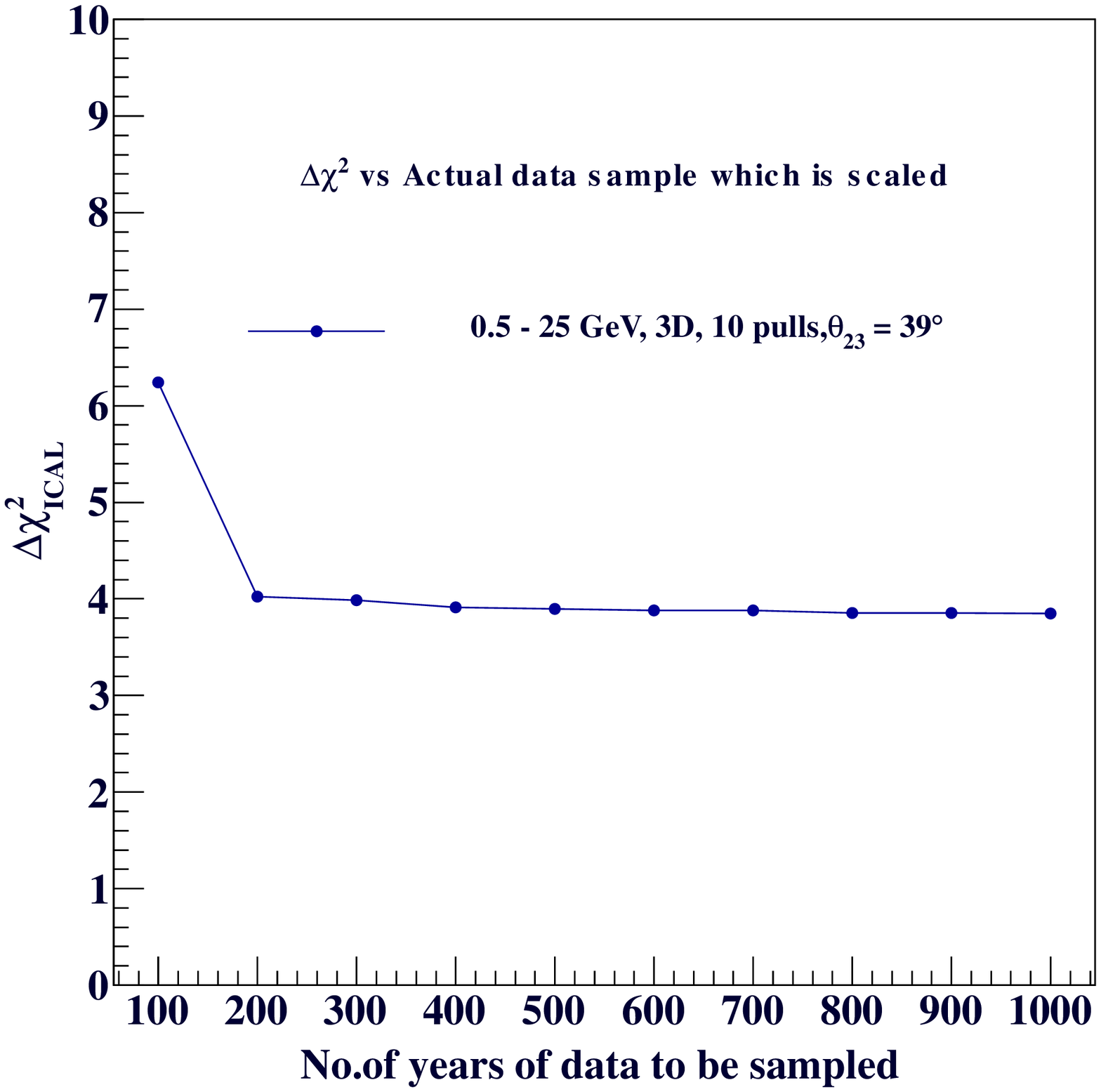}
\hfill
\includegraphics[width=0.45\textwidth]{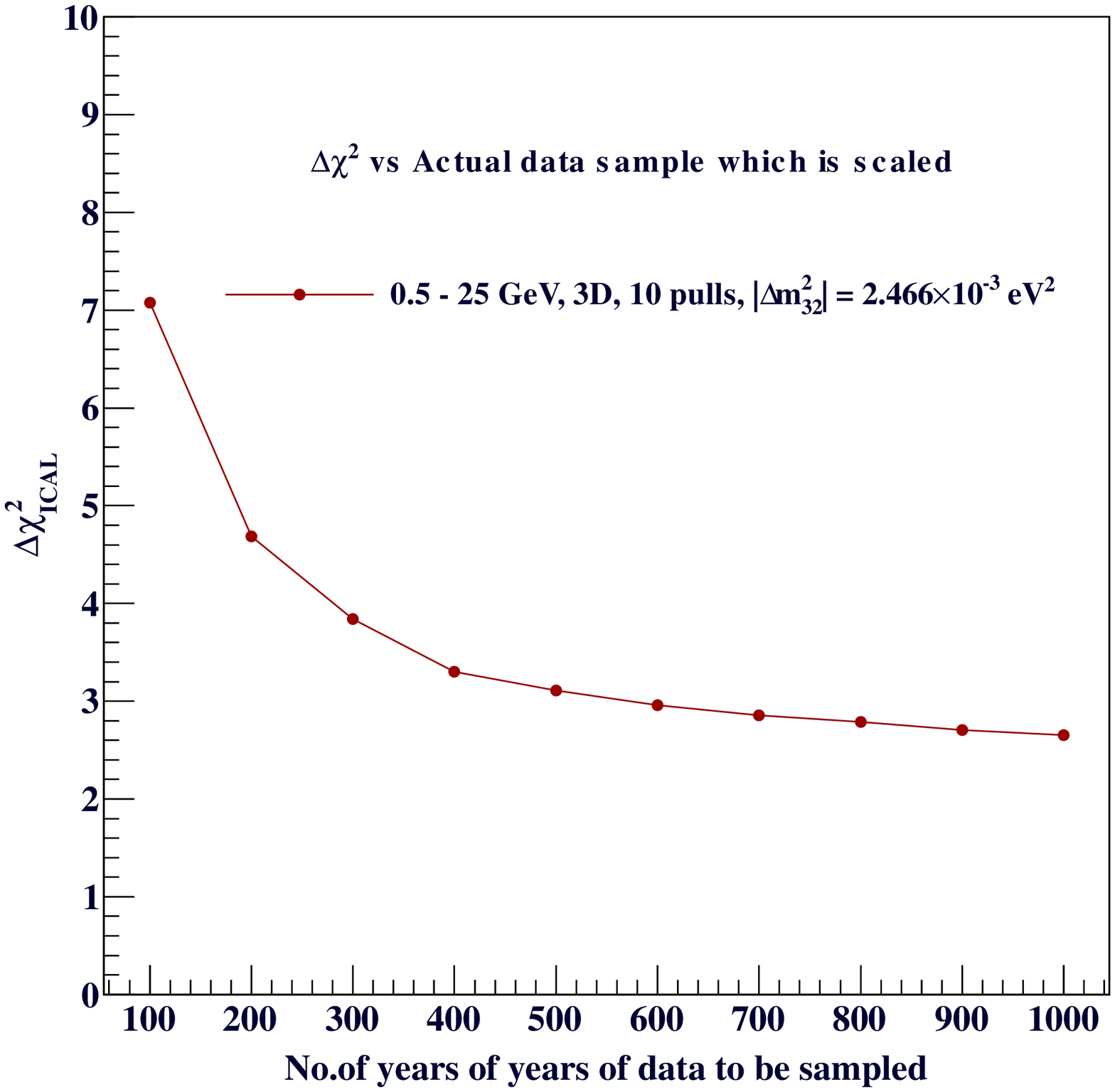}
\caption{The value of $\Delta\chi^2_{ICAL}$ vs the number of years
of sample theory to be scaled to 10 years for $\theta_{23}=39^\circ$
(left) and $|\Delta{m}_{32}^2| = 2.466 \times 10^{-3} {\rm eV}^2$
(right). It can be seen that the scaling of only 100 years of sample
size gives too high a $\chi^2$ which is more prominent in the case of
$\Delta m^2_{32}$. As the number of years of sample to be scaled
increases, the value of $\chi^2$ also comes down as well as saturates
for large enough sample sizes.}
\label{chi2-fluct}
\end{figure}

As discussed in Section~\ref{nevt}, the analyses in the previous
sections were done by taking a 1000 year sample of charged current muon
neutrino events and scaling it down to the required number of years
for comparison with ``data''. It is important to reduce ``theory''
fluctuations so as to obtain a genuine result from the analysis. The
effect of taking different years of exposure as theory samples, to
be scaled to 10 years on precision measurements of $\theta_{23}$ and
$|\Delta{m^2_{32}}|$, is shown in Fig.~\ref{chi2-fluct} for arbitrarily
chosen values of $\theta_{23} = 39^\circ$ and $\Delta m^2_{32} =
2.466\times 10^{-3}$ eV$^2$ ($\Delta m^2_{\rm eff} = 2.5\times 10^{-3}$
eV$^2$). It can be seen that a smaller sample has more fluctuations and
hence yields a larger value of $\Delta\chi^2_{ICAL}$ for a given parameter
thus giving too good a precision on the oscillation parameter which is
false. The larger sample takes care of this by reducing the ($\sqrt{N}$)
fluctuations in the theory itself. The $\chi^2$ stabilises at a point when
the sample is fairly large and this is the number of years of exposure
to be taken and scaled down for the analysis. It can be seen that use of
1000 year sample yields fairly stable results and thus the analyses have
made use of such a sample of charged current muon neutrino events. We
note however that the results are much more sensitive to the sample size
in the case of $\vert \Delta m^2_{32} \vert$ than for $\theta_{23}$.

\section*{Acknowledgments}
We thank Prof.~Nita Sinha, IMSc, for many discussions. We also thank
Meghna K.K. for the muon resolution table in the lower muon energy
range. We thank the INO internal referees for their valuable and
insightful comments on the draft. LSM thanks DAE India and DST India,
for funding this research and is thankful for the excellent computing
facilities and support from the computing section of IMSc, Chennai,
which made the extensive computing for this analysis possible.

\end{document}